\documentclass[twocolumn]{IEEEtran}
\usepackage{amsmath}
\usepackage{amssymb}
\usepackage{amsfonts}
\usepackage{slashbox}
\usepackage{ifpdf}
\ifpdf
	\usepackage[pdftex]{graphicx}
  \usepackage[update]{epstopdf}
\else
	\usepackage{graphicx}
\fi
\usepackage{cite, amsmath, amsfonts, amssymb, psfrag, epsfig}
\newtheorem{definition}{{Definition}}
\newtheorem{theorem}{{Theorem}}
\newtheorem{lemma}{{Lemma}}

\newtheorem{corollary}{{Corollary}}

\DeclareMathAlphabet{\mathpzc}{OT1}{pzc}{m}{it}

\begin{document}

\title{Broadcasting Correlated Vector Gaussians}


\author{Lin Song, Jun Chen, and Chao Tian
\thanks{L. Song is with the Institute of Network Coding, The Chinese Univeristy of Hong Kong, Shatin, Hong Kong (email: linsong@inc.cuhk.edu.hk).}
\thanks{J. Chen is with the Department
of Electrical and Computer Engineering, McMaster University,
Hamilton, ON L8S 4K1, Canada  (email: junchen@ece.mcmaster.ca).}
\thanks{C. Tian is with the University of Tennessee, Knoxville, TN 37996, USA (email: chao.tian@utk.edu).}}

\maketitle

\begin{abstract}
The problem of sending two correlated vector Gaussian sources over a bandwidth-matched two-user scalar Gaussian broadcast channel is studied in this work, where each receiver wishes to reconstruct its target source under a covariance distortion constraint. We derive a lower bound on the optimal tradeoff between the transmit  power and the achievable reconstruction distortion pair. Our derivation is based on a new bounding technique which involves the introduction of appropriate remote sources. Furthermore, it is shown that this lower bound is achievable by a class of hybrid schemes for the special case where the weak receiver wishes to reconstruct a  scalar source under the mean squared error distortion constraint.
\end{abstract}


\section{Introduction}\label{sec:introduction}

Unlike in point-to-point communication systems where the source-channel separation architecture is optimal \cite{Shannon:48}, in multi-user systems, a separation-based architecture is usually suboptimal.
In such scenarios, hybrid schemes have emerged as a promising approach to gain performance improvement over either pure digital schemes (separation-based schemes) or pure analog schemes, e.g., in \cite{MittalPhamdo:02} for bandwidth-mismatch Gaussian source broadcast (see also \cite{PPR11,BAL11,KKE12} for variants of this problem), and in \cite{Lapidoth:09} for sending a bivariate Gaussian source over a Gaussian multiple access channel. Recently,  building upon the important work by Bross \textit{et al.} \cite{BLT10} as well as \cite{SV09} and \cite{GT13},   Tian \textit{et al.}  \cite{TDS11} showed that, for the problem of broadcasting a bivariate Gaussian source,  hybrid schemes are not only able to provide such performance improvement, they can in fact be optimal.

\begin{figure}[tb]
\begin{centering}
\includegraphics[width=9.5cm]{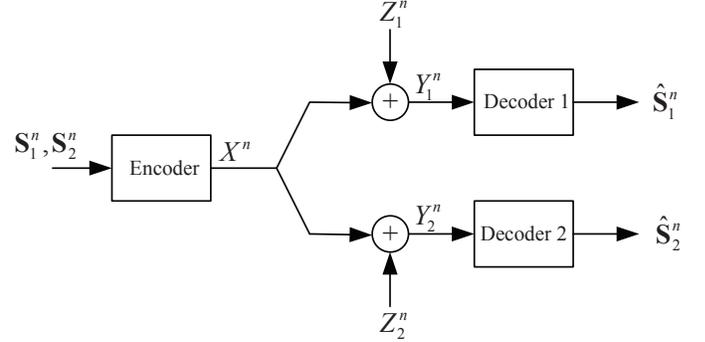}
\caption{Broadcasting correlated vector Gaussian sources.\label{fig:system}}
\end{centering}
\end{figure}

In this paper, we consider the problem of sending two correlated vector Gaussian sources over a bandwidth-matched two-user scalar Gaussian broadcast channel, where each receiver wishes to reconstruct its target source under a covariance distortion constraint (see Fig. \ref{fig:system}). This can be viewed as a vector generalization of the problem studied in \cite{BLT10,SV09,TDS11}. We derive a lower bound on the optimal tradeoff between the transmit  power and the achievable reconstruction distortion pair. Furthermore, it is shown that this lower bound is tight for the scenario, referred to as the vector-scalar case, where the weak receiver wishes to reconstruct a  scalar source under the mean squared error distortion constraint. It is worth noting that  the brute-force proof method in \cite{BLT10, TDS11} is difficult to generalize to the problem being considered. Therefore, instead of seeking explicit upper and lower bounds and showing their tightness by direct comparison, we take a more conceptual approach in the present work. In particular, the derivation of our lower bound is based on a new bounding technique which involves the introduction of appropriate remote sources; moreover, to obtain a matching upper bound in the vector-scalar case, we construct a scheme with its parameters specified according to an optimization problem motivated by the lower bound. Another finding is that the optimal scheme is in general not unique. Indeed, we show that, in the vector-scalar case, the optimal tradeoff between the transmit power and the reconstruction distortion pair is achievable by a class of hybrid schemes, which includes the scheme proposed by Tian \textit{et al.}  \cite{TDS11} as an extremal example.

\section{Problem Definition}\label{sec:definition}

Let $\mathbf{S}_i$ be an $m_i\times 1$ zero-mean random vector, $i=1, 2$. We assume that $\mathbf{S}_1$ and $\mathbf{S}_2$ are jointly Gaussian with covariance matrix
\begin{align*}
\mathbf{\Sigma}_{\mathbf{S}_1,\mathbf{S}_2}=\left(
                                              \begin{array}{cc}
                                                \mathbf{\Sigma}_{\mathbf{S}_1} & \mathbb{E}[\mathbf{S}_1\mathbf{S}^T_2] \\
                                                \mathbb{E}[\mathbf{S}_2\mathbf{S}^T_1] & \mathbf{\Sigma}_{\mathbf{S}_2} \\
                                              \end{array}
                                            \right),
\end{align*}
where $\mathbf{\Sigma}_{\mathbf{S}_i}=\mathbb{E}[\mathbf{S}_i\mathbf{S}^T_i]$, $i=1,2$. Let the broadcast channel additive noises $Z_1$ and $Z_2$ be two zero-mean Gaussian random variables, jointly independent of $(\mathbf{S}_1,\mathbf{S}_2)$, with variances $N_1$ and $N_2$, respectively; it is assumed that $N_2>N_1$. Let $\{(\mathbf{S}_1(t),\mathbf{S}_2(t),Z_1(t),Z_2(t))\}_{t=1}^\infty$ be i.i.d. copies of $(\mathbf{S}_1,\mathbf{S}_2,Z_1,Z_2)$.

\begin{definition}
An $(n,P,\mathbf{D}_1,\mathbf{D}_2)$ source-channel broadcast code consists of an encoding function $f:\mathbb{R}^{m_1\times n}\times\mathbb{R}^{m_2\times n}\rightarrow\mathbb{R}^n$ and two decoding function $g_i:\mathbb{R}^n\rightarrow\mathbb{R}^{m_i\times n}$, $i=1,2$, such that
\begin{align*}
&\frac{1}{n}\mathbb{E}[X^n(X^n)^T]\leq P,\\
&\frac{1}{n}\mathbb{E}[(\mathbf{S}^n_i-\hat{\mathbf{S}}^n_i)(\mathbf{S}^n_i-\hat{\mathbf{S}}^n_i)^T]\preceq\mathbf{D}_i,\quad i=1,2,
\end{align*}
where $X^n= f(\mathbf{S}^n_1,\mathbf{S}^n_2)$ and $\hat{\mathbf{S}}^n_i= g_i(Y^n_i)$, $i=1,2$, with $Y^n_i=X^n+Z^n_i$, $i=1,2$.
\end{definition}

It is clear that the performance of any source-channel broadcast code depends on $(Z^n_1,Z^n_2)$ only through their marginal distributions. Therefore, we shall assume the broadcast channel is physically degraded and write $Z^n_2=Z^n_1+Z^n$, where $Z^n$ is a zero-mean Gaussian random vector with i.i.d. entries of variance $N_2-N_1$ and is independent of $Z^n_1$. It is also clear \cite[App. 3.A]{KSH00} that there is no loss of optimality in assuming $\hat{\mathbf{S}}^n_i=g_i(Y^n_i)=\mathbb{E}[\mathbf{S}^n_i|Y^n_i]$, $i=1,2$.

\begin{definition}
We say $(P,\mathbf{D}_1,\mathbf{D}_2)$ is achievable if there exists an $(n,P,\mathbf{D}_1,\mathbf{D}_2)$ source-channel broadcast code. Let $\mathcal{PD}$ denote the closure of the set of all achievable $(P,\mathbf{D}_1,\mathbf{D}_2)$.
\end{definition}

\begin{definition}
Let $P(\mathbf{D}_1,\mathbf{D}_2)=\inf\{P:(P,\mathbf{D}_1,\mathbf{D}_2)\in\mathcal{PD}\}$.
\end{definition}

With the above definitions, it is clear that the fundamental problem in this joint source-channel coding scenario is to determine the function $P(\mathbf{D}_1,\mathbf{D}_2)$, which characterizes the optimal tradeoff between the transmit  power and the achievable reconstruction distortion pair\footnote{This formulation is slightly different from that in \cite{BLT10, TDS11}, where the power $P$ is fixed, and the tradeoff between the reconstruction distortion pair is considered. We find the current formulation more suitable here, since both receivers are to reconstruct vector sources.}. Unless specified otherwise, we assume $\mathbf{\Sigma}_{\mathbf{S}_1,\mathbf{S}_2}\succ\mathbf{0}$ and $\mathbf{D}_i\succ\mathbf{0}$, $i=1,2$.

The remainder of this paper is organized as follows. We derive a lower bound on $P(\mathbf{D}_1,\mathbf{D}_2)$ in Section \ref{sec:lower}. It is shown in Section \ref{sec:upper} that, for the vector-scalar case, this lower bound is achievable by a class of hybrid schemes. We conclude the paper in Section \ref{sec:conclusion}. Throughout this paper, the logarithm function is to base $e$.

\section{Lower Bound}\label{sec:lower}

Let $\mathbf{U}_i$ be an $m_i\times 1$ zero-mean random vector, $i=1, 2$. We assume that $\mathbf{U}_1$ and $\mathbf{U}_2$ are jointly Gaussian with covariance matrix
\begin{align*}
\mathbf{\Sigma}_{\mathbf{U}_1,\mathbf{U}_2}=\left(
                                              \begin{array}{cc}
                                                \mathbf{\Sigma}_{\mathbf{U}_1} & \mathbb{E}[\mathbf{U}_1\mathbf{U}^T_2] \\
                                                \mathbb{E}[\mathbf{U}_2\mathbf{U}^T_1] & \mathbf{\Sigma}_{\mathbf{U}_2} \\
                                              \end{array}
                                            \right),
\end{align*}
where $\mathbf{\Sigma}_{\mathbf{U}_i}=\mathbb{E}[\mathbf{U}_i\mathbf{U}^T_i]$, $i=1,2$.

The main result of this section is the following theorem.
\begin{theorem}\label{thm:lower}
\begin{align}
P(\mathbf{D}_1,\mathbf{D}_2)&\geq\inf\limits_{\mathbf{\Theta}}\sup\limits_{\mathbf{\Sigma}_{\mathbf{U}_1,\mathbf{U}_2}\succ \mathbf{0}}N_1\frac{|\mathbf{\Sigma}_{\mathbf{S}_1,\mathbf{S}_2}||\mathbf{\Theta}_2+\mathbf{\Sigma}_{\mathbf{U}_2}|}{|\mathbf{\Theta}||\mathbf{D}_2+\mathbf{\Sigma}_{\mathbf{U}_2}|}\nonumber\\
&\qquad+(N_2-N_1)\frac{|\mathbf{\Sigma}_{\mathbf{S}_2}+\mathbf{\Sigma}_{\mathbf{U}_2}|}{|\mathbf{D}_2+\mathbf{\Sigma}_{\mathbf{U}_2}|}-N_2\label{eq:objective}
\end{align}
with the infimum taken over $(m_1+m_2)\times (m_1+m_2)$ matrix $\mathbf{\Theta}$
subject to the constraints
\begin{align}
&\mathbf{0}\prec\mathbf{\Theta}\preceq\mathbf{\Sigma}_{\mathbf{S}_1,\mathbf{S}_2},\label{eq:thetacont1}\\
&\mathbf{\Theta}_1\preceq\mathbf{D}_1.\label{eq:thetacont2}
\end{align}
Here we assume that $\mathbf{\Theta}$ is partitioned to the form
\begin{align*}
\mathbf{\Theta}=\left(
                  \begin{array}{cc}
                    \mathbf{\Theta}_1 & \# \\
                    \# & \mathbf{\Theta}_2 \\
                  \end{array}
                \right),
\end{align*}
where $ \mathbf{\Theta}_i$ is of size $m_i\times m_i$ for $i=1,2$.
\end{theorem}

\textit{Remark:} It is interesting to note that the objective function on the right-hand side of (\ref{eq:objective}) depends on $\mathbf{\Sigma}_{\mathbf{U}_1,\mathbf{U}_2}$ only through $\mathbf{\Sigma}_{\mathbf{U}_2}$. Therefore, one can simply take the supremum in (\ref{eq:objective}) over $\mathbf{\Sigma}_{\mathbf{U}_2}\succ\mathbf{0}$.

\textit{Remark:} Theorem \ref{thm:lower} is in fact closely related to \cite[Th. 1]{RFZ06}. A detailed explanation of the connections between these two results can be found in \cite{KC14}.

The following two elementary inequalities are needed for the proof of Theorem \ref{thm:lower}. For completeness, their proofs are given in Appendices \ref{app:entropybound} and \ref{app:condentropy}.

\begin{lemma}\label{lem:entropybound}
For any $m\times n$ random matrices $\mathbf{W}$ and $\hat{\mathbf{W}}$,
\begin{align*}
h(\mathbf{W}|\hat{\mathbf{W}})\leq\frac{n}{2}\log\left|\frac{2\pi e}{n}\mathbb{E}[(\mathbf{W}-\hat{\mathbf{W}})(\mathbf{W}-\hat{\mathbf{W}})^T]\right|.
\end{align*}
\end{lemma}


\begin{lemma}\label{lem:condentropy}
Let $\mathbf{W}_i$ be an $m_i\times n$ zero-mean random matrix, $i=1,2$. If $\frac{1}{n}\mathbb{E}[(\mathbf{W}^T_1,\mathbf{W}^T_2)^T(\mathbf{W}^T_1,\mathbf{W}^T_2)]\succ\mathbf{0}$, then
\begin{align*}
h(\mathbf{W}_1|\mathbf{W}_2)\leq\frac{n}{2}\log\frac{\left|\frac{2\pi e}{n}\mathbb{E}[(\mathbf{W}^T_1,\mathbf{W}^T_2)^T(\mathbf{W}^T_1,\mathbf{W}^T_2)]\right|}{\left|\frac{2\pi e}{n}\mathbb{E}[\mathbf{W}_2\mathbf{W}^T_2]\right|}.
\end{align*}
\end{lemma}

Now we proceed to prove Theorem \ref{thm:lower}.

\begin{IEEEproof}[Proof of Theorem \ref{thm:lower}]
For any $(n,P,\mathbf{D}_1,\mathbf{D}_2)$ source-channel broadcast code, let  $\hat{\mathbf{S}}^n_i=g_i(Y^n_i)=\mathbb{E}[\mathbf{S}^n_i|Y^n_i]$, $i=1,2$, and $\tilde{\mathbf{S}}^n_2=\mathbb{E}[\mathbf{S}^n_2|Y^n_1]$; furthermore, let
\begin{align*}
\mathbf{\Theta}=\left(
                  \begin{array}{cc}
                    \mathbf{\Theta}_1 & \mathbf{\Upsilon} \\
                    \mathbf{\Upsilon}^T & \mathbf{\Theta}_2 \\
                  \end{array}
                \right)
\end{align*}
with $\mathbf{\Theta}_1=\frac{1}{n}\mathbb{E}[(\mathbf{S}^n_1-\hat{\mathbf{S}}^n_1)(\mathbf{S}^n_1-\hat{\mathbf{S}}^n_1)^T]$, $\mathbf{\Theta}_2=\frac{1}{n}\mathbb{E}[(\mathbf{S}^n_2-\tilde{\mathbf{S}}^n_2)(\mathbf{S}^n_2-\tilde{\mathbf{S}}^n_2)^T]$, and $\mathbf{\Upsilon}=\frac{1}{n}\mathbb{E}[(\mathbf{S}^n_1-\hat{\mathbf{S}}^n_1)(\mathbf{S}^n_2-\tilde{\mathbf{S}}^n_2)^T]$. Note that $\mathbf{\Theta}$ satisfies (\ref{eq:thetacont1}) and (\ref{eq:thetacont2}). Therefore, it suffices to show that
\begin{align}
P&\geq N_1\frac{|\mathbf{\Sigma}_{\mathbf{S}_1,\mathbf{S}_2}||\mathbf{\Theta}_2+\mathbf{\Sigma}_{\mathbf{U}_2}|}{|\mathbf{\Theta}||\mathbf{D}_2+\mathbf{\Sigma}_{\mathbf{U}_2}|}\nonumber\\
&\quad+(N_2-N_1)\frac{|\mathbf{\Sigma}_{\mathbf{S}_2}+\mathbf{\Sigma}_{\mathbf{U}_2}|}{|\mathbf{D}_2+\mathbf{\Sigma}_{\mathbf{U}_2}|}-N_2\label{eq:proof1}
\end{align}
for all $\mathbf{\Sigma}_{\mathbf{U}_1,\mathbf{U}_2}\succ\mathbf{0}$.

Let $\{\mathbf{U}_1(t),\mathbf{U}_2(t)\}_{t=1}^n$ be i.i.d. copies of $(\mathbf{U}_1,\mathbf{U}_2)$. We assume that $(\mathbf{U}^n_1,\mathbf{U}^n_2)$ is independent of $(\mathbf{S}^n_1,\mathbf{S}^n_2,Z^n_1,Z^n)$.
Define $\mathbf{V}^n_i=\mathbf{S}^n_i+\mathbf{U}^n_i$, $i=1,2$. Here $\mathbf{V}_1$ and $\mathbf{V}_2$ can be understood as the remote sources that should be reconstructed, yet the encoder only has access to $\mathbf{S}_1$ and  $\mathbf{S}_2$. The introduction of $(\mathbf{V}_1,\mathbf{V}_2)$ is partly inspired by Ozarow's converse argument for the Gaussian multiple description problem \cite{Ozarow80} (see also \cite{WV07,Chen09,SSC13}).

We shall first bound $I(\mathbf{V}^n_2;Y^n_2)$. In view of the fact that
\begin{align*}
0\leq I(\mathbf{V}^n_2;Y^n_2)\leq I(X^n;Y^n_2)\leq\frac{n}{2}\log\frac{P+N_2}{N_2},
\end{align*}
we have
\begin{align}
I(\mathbf{V}^n_2;Y^n_2)=\frac{n}{2}\log\frac{P+N_2}{\alpha P+N_2}\label{eq:comb2}
\end{align}
for some $\alpha\in[0,1]$. On the other hand,
\begin{align}
&I(\mathbf{V}^n_2;Y^n_2)\nonumber\\
&=h(\mathbf{V}^n_2)-h(\mathbf{V}^n_2|Y^n_2)\nonumber\\
&=\frac{n}{2}\log|2\pi e(\mathbf{\Sigma}_{\mathbf{S}_2}+\mathbf{\Sigma}_{\mathbf{U}_2})|-h(\mathbf{V}^n_2|Y^n_2)\nonumber\\
&\geq\frac{n}{2}\log|2\pi e(\mathbf{\Sigma}_{\mathbf{S}_2}+\mathbf{\Sigma}_{\mathbf{U}_2})|-h(\mathbf{V}^n_2|\hat{\mathbf{S}}^n_2)\nonumber\\
&\geq\frac{n}{2}\log|\mathbf{\Sigma}_{\mathbf{S}_2}+\mathbf{\Sigma}_{\mathbf{U}_2}|\nonumber\\
&\quad-\frac{n}{2}\log\left|\frac{1}{n}\mathbb{E}[(\mathbf{V}^n_2-\hat{\mathbf{S}}^n_2)(\mathbf{V}^n_2-\hat{\mathbf{S}}^n_2)^T]\right|\label{eq:entropybound}\\
&\geq\frac{n}{2}\log|\mathbf{\Sigma}_{\mathbf{S}_2}+\mathbf{\Sigma}_{\mathbf{U}_2}|\nonumber\\
&\quad-\frac{n}{2}\log\left|\frac{1}{n}\mathbb{E}[(\mathbf{S}^n_2-\hat{\mathbf{S}}^n_2)(\mathbf{S}^n_2-\hat{\mathbf{S}}^n_2)^T]+\mathbf{\Sigma}_{\mathbf{U}_2}\right|\nonumber\\
&\geq\frac{n}{2}\log\frac{|\mathbf{\Sigma}_{\mathbf{S}_2}+\mathbf{\Sigma}_{\mathbf{U}_2}|}{|\mathbf{D}_2+\mathbf{\Sigma}_{\mathbf{U}_2}|},\label{eq:comb1}
\end{align}
where (\ref{eq:entropybound}) follows from Lemma \ref{lem:entropybound}.
 Combining (\ref{eq:comb2}) and (\ref{eq:comb1}) gives
\begin{align}
\frac{|\mathbf{\Sigma}_{\mathbf{S}_2}+\mathbf{\Sigma}_{\mathbf{U}_2}|}{|\mathbf{D}_2+\mathbf{\Sigma}_{\mathbf{U}_2}|}\leq\frac{P+N_2}{\alpha P+N_2}.\label{eq:main1}
\end{align}
Now we proceed to bound $I(\mathbf{V}^n_1;Y^n_1|\mathbf{V}^n_2)$.
Since $h(Y^n_2)\leq\frac{n}{2}\log(2\pi e(P+N_2))$, it follows from (\ref{eq:comb2}) that
\begin{align}
h(Y^n_2|\mathbf{V}^n_2)\leq\frac{n}{2}\log(2\pi e(\alpha P+N_2)).\label{eq:imply}
\end{align}
By the entropy power inequality,
\begin{align*}
h(Y^n_2|\mathbf{V}^n_2)&\geq\frac{n}{2}\log\Big(e^{\frac{2}{n}h(Y^n_1|\mathbf{V}^n_2)}+e^{\frac{2}{n}h(Z^n)}\Big)\\
&=\frac{n}{2}\log\Big(e^{\frac{2}{n}h(Y^n_1|\mathbf{V}^n_2)}+2\pi e(N_2-N_1)\Big),
\end{align*}
which, together with (\ref{eq:imply}), implies
\begin{align*}
h(Y^n_1|\mathbf{V}^n_2)\leq\frac{n}{2}\log(2\pi e(\alpha P+N_1)).
\end{align*}
Note that
\begin{align}
&I(\mathbf{V}^n_1;Y^n_1|\mathbf{V}^n_2)\nonumber\\
&=h(Y^n_1|\mathbf{V}^n_2)-h(Y^n_1|\mathbf{V}^n_1,\mathbf{V}^n_2)\nonumber\\
&\leq\frac{n}{2}\log(2\pi e(\alpha P+N_1))-h(Y^n_1|\mathbf{V}^n_1,\mathbf{V}^n_2)\nonumber\\
&=\frac{n}{2}\log\frac{\alpha P+N_1}{N_1}-h(Y^n_1|\mathbf{V}^n_1,\mathbf{V}^n_2)+h(Y^n_1|\mathbf{S}^n_1,\mathbf{S}^n_2)\nonumber\\
&=\frac{n}{2}\log\frac{\alpha P+N_1}{N_1}-I(\mathbf{S}^n_1,\mathbf{S}^n_2;Y^n_1|\mathbf{V}^n_1,\mathbf{V}^n_2)\nonumber\\
&=\frac{n}{2}\log\frac{\alpha P+N_1}{N_1}-\frac{n}{2}\log\frac{|2\pi e\mathbf{\Sigma}_{\mathbf{S}_1,\mathbf{S}_2}||2\pi e\mathbf{\Sigma}_{\mathbf{U}_1,\mathbf{U}_2}|}{|2\pi e(\mathbf{\Sigma}_{\mathbf{S}_1,\mathbf{S}_2}+\mathbf{\Sigma}_{\mathbf{U}_1,\mathbf{U}_2})|}\nonumber\\
&\quad+h(\mathbf{S}^n_1,\mathbf{S}^n_2|\mathbf{V}^n_1,\mathbf{V}^n_2,Y^n_1)\nonumber\\
&=\frac{n}{2}\log\frac{\alpha P+N_1}{N_1}-\frac{n}{2}\log\frac{|2\pi e\mathbf{\Sigma}_{\mathbf{S}_1,\mathbf{S}_2}||2\pi e\mathbf{\Sigma}_{\mathbf{U}_1,\mathbf{U}_2}|}{|2\pi e(\mathbf{\Sigma}_{\mathbf{S}_1,\mathbf{S}_2}+\mathbf{\Sigma}_{\mathbf{U}_1,\mathbf{U}_2})|}\nonumber\\
&\quad+h(\mathbf{S}^n_1-\hat{\mathbf{S}}^n_1,\mathbf{S}^n_2-\tilde{\mathbf{S}}^n_2|\mathbf{V}^n_1-\hat{\mathbf{S}}^n_1,\mathbf{V}^n_2-\tilde{\mathbf{S}}^n_2,Y^n_1)\nonumber\\
&\leq\frac{n}{2}\log\frac{\alpha P+N_1}{N_1}-\frac{n}{2}\log\frac{|2\pi e\mathbf{\Sigma}_{\mathbf{S}_1,\mathbf{S}_2}||2\pi e\mathbf{\Sigma}_{\mathbf{U}_1,\mathbf{U}_2}|}{|2\pi e(\mathbf{\Sigma}_{\mathbf{S}_1,\mathbf{S}_2}+\mathbf{\Sigma}_{\mathbf{U}_1,\mathbf{U}_2})|}\nonumber\\
&\quad+h(\mathbf{S}^n_1-\hat{\mathbf{S}}^n_1,\mathbf{S}^n_2-\tilde{\mathbf{S}}^n_2|\mathbf{V}^n_1-\hat{\mathbf{S}}^n_1,\mathbf{V}^n_2-\tilde{\mathbf{S}}^n_2)\nonumber\\
&\leq\frac{n}{2}\log\frac{\alpha P+N_1}{N_1}-\frac{n}{2}\log\frac{|\mathbf{\Sigma}_{\mathbf{S}_1,\mathbf{S}_2}||\mathbf{\Sigma}_{\mathbf{U}_1,\mathbf{U}_2}|}{|\mathbf{\Sigma}_{\mathbf{S}_1,\mathbf{S}_2}+\mathbf{\Sigma}_{\mathbf{U}_1,\mathbf{U}_2}|}\nonumber\\
&\quad+\frac{n}{2}\log\frac{|\mathbf{\Theta}||\mathbf{\Sigma}_{\mathbf{U}_1,\mathbf{U}_2}|}{|\mathbf{\Theta}+\mathbf{\Sigma}_{\mathbf{U}_1,\mathbf{U}_2}|}\label{eq:worstnoise}\\
&=\frac{n}{2}\log\frac{(\alpha P+N_1)|\mathbf{\Sigma}_{\mathbf{S}_1,\mathbf{S}_2}+\mathbf{\Sigma}_{\mathbf{U}_1,\mathbf{U}_2}||\mathbf{\Theta}|}{N_1|\mathbf{\Sigma}_{\mathbf{S}_1,\mathbf{S}_2}||\mathbf{\Theta}+\mathbf{\Sigma}_{\mathbf{U}_1,\mathbf{U}_2}|},\label{eq:comb3}
\end{align}
where (\ref{eq:worstnoise}) is due to Lemma \ref{lem:condentropy}.
On the other hand,
\begin{align}
&I(\mathbf{V}^n_1;Y^n_1|\mathbf{V}^n_2)\nonumber\\
&=h(\mathbf{V}^n_1|\mathbf{V}^n_2)-h(\mathbf{V}^n_1|\mathbf{V}^n_2,Y^n_1)\nonumber\\
&=\frac{n}{2}\log\frac{|2\pi e(\mathbf{\Sigma}_{\mathbf{S}_1,\mathbf{S}_2}+\mathbf{\Sigma}_{\mathbf{U}_1,\mathbf{U}_2})|}{|2\pi e(\mathbf{\Sigma}_{\mathbf{S}_2}+\mathbf{\Sigma}_{\mathbf{U}_2})|}-h(\mathbf{V}^n_1|\mathbf{V}^n_2,Y^n_1)\nonumber\\
&=\frac{n}{2}\log\frac{|2\pi e(\mathbf{\Sigma}_{\mathbf{S}_1,\mathbf{S}_2}+\mathbf{\Sigma}_{\mathbf{U}_1,\mathbf{U}_2})|}{|2\pi e(\mathbf{\Sigma}_{\mathbf{S}_2}+\mathbf{\Sigma}_{\mathbf{U}_2})|}\nonumber\\
&\quad-h(\mathbf{V}^n_1-\hat{\mathbf{S}}^n_1|\mathbf{V}^n_2-\tilde{\mathbf{S}}^n_2,Y^n_1)\nonumber\\
&\geq\frac{n}{2}\log\frac{|2\pi e(\mathbf{\Sigma}_{\mathbf{S}_1,\mathbf{S}_2}+\mathbf{\Sigma}_{\mathbf{U}_1,\mathbf{U}_2})|}{|2\pi e(\mathbf{\Sigma}_{\mathbf{S}_2}+\mathbf{\Sigma}_{\mathbf{U}_2})|}-h(\mathbf{V}^n_1-\hat{\mathbf{S}}^n_1|\mathbf{V}^n_2-\tilde{\mathbf{S}}^n_2)\nonumber\\
&\geq\frac{n}{2}\log\frac{|2\pi e(\mathbf{\Sigma}_{\mathbf{S}_1,\mathbf{S}_2}+\mathbf{\Sigma}_{\mathbf{U}_1,\mathbf{U}_2})|}{|2\pi e(\mathbf{\Sigma}_{\mathbf{S}_2}+\mathbf{\Sigma}_{\mathbf{U}_2})|}\nonumber\\
&\quad-\frac{n}{2}\log\frac{|2\pi e(\mathbf{\Theta}+\mathbf{\Sigma}_{\mathbf{U}_1,\mathbf{U}_2})|}{|2\pi e(\mathbf{\Theta}_2+\mathbf{\Sigma}_{\mathbf{U}_2})|}\label{eq:entropybound2}\\
&=\frac{n}{2}\log\frac{|\mathbf{\Sigma}_{\mathbf{S}_1,\mathbf{S}_2}+\mathbf{\Sigma}_{\mathbf{U}_1,\mathbf{U}_2}||\mathbf{\Theta}_2+\mathbf{\Sigma}_{\mathbf{U}_2}|}{|\mathbf{\Sigma}_{\mathbf{S}_2}+\mathbf{\Sigma}_{\mathbf{U}_2}||\mathbf{\Theta}+\mathbf{\Sigma}_{\mathbf{U}_1,\mathbf{U}_2}|},\label{eq:comb4}
\end{align}
where (\ref{eq:entropybound2}) follows from Lemma \ref{lem:condentropy}.
Combining (\ref{eq:comb3}) and (\ref{eq:comb4}) yields
\begin{align}
\frac{|\mathbf{\Theta}_2+\mathbf{\Sigma}_{\mathbf{U}_2}|}{|\mathbf{\Sigma}_{\mathbf{S}_2}+\mathbf{\Sigma}_{\mathbf{U}_2}|}\leq\frac{(\alpha P+N_1)|\mathbf{\Theta}|}{N_1|\mathbf{\Sigma}_{\mathbf{S}_1,\mathbf{S}_2}|}.\label{eq:main2}
\end{align}
One can readily obtain (\ref{eq:proof1}) from (\ref{eq:main1}) and (\ref{eq:main2})  by eliminating $\alpha$.
This completes the proof of Theorem \ref{thm:lower}.
\end{IEEEproof}

This theorem leads us to the following (potentially weakened) lower bound on $P(\mathbf{D}_1,\mathbf{D}_2)$. Somewhat surprisingly, this lower bound turns out to be tight in the vector-scalar case.

\begin{corollary}\label{cor:lowersimplified}
\begin{align*}
P(\mathbf{D}_1,\mathbf{D}_2)&\geq\sup\limits_{\mathbf{\Sigma}_{\mathbf{U}_1,\mathbf{U}_2}\succ \mathbf{0}}N_1\frac{|\mathbf{\Sigma}_{\mathbf{S}_1,\mathbf{S}_2}+\mathbf{\Sigma}_{\mathbf{U}_1,\mathbf{U}_2}|}{|\mathbf{D}_1+\mathbf{\Sigma}_{\mathbf{U}_1}||\mathbf{D}_2+\mathbf{\Sigma}_{\mathbf{U}_2}|}\\
&\hspace{0.7in}+(N_2-N_1)\frac{|\mathbf{\Sigma}_{\mathbf{S}_2}+\mathbf{\Sigma}_{\mathbf{U}_2}|}{|\mathbf{D}_2+\mathbf{\Sigma}_{\mathbf{U}_2}|}-N_2.
\end{align*}
\end{corollary}

\begin{IEEEproof}[Proof of Corollary \ref{cor:lowersimplified}]
Note that
\begin{align}
&\frac{|\mathbf{\Sigma}_{\mathbf{S}_1,\mathbf{S}_2}||\mathbf{\Theta}_2+\mathbf{\Sigma}_{\mathbf{U}_2}|}{|\mathbf{\Theta}||\mathbf{D}_2+\mathbf{\Sigma}_{\mathbf{U}_2}|}\nonumber\\
&=\frac{|\mathbf{\Sigma}_{\mathbf{S}_1,\mathbf{S}_2}||\mathbf{\Theta}_2+\mathbf{\Sigma}_{\mathbf{U}_2}||\mathbf{\Theta}+\mathbf{\Sigma}_{\mathbf{U}_1,\mathbf{U}_2}|}{|\mathbf{\Theta}||\mathbf{D}_2+\mathbf{\Sigma}_{\mathbf{U}_2}||\mathbf{\Theta}+\mathbf{\Sigma}_{\mathbf{U}_1,\mathbf{U}_2}|}.\label{eq:identity}
\end{align}
For any $\mathbf{\Theta}$ satisfying (\ref{eq:thetacont1}) and (\ref{eq:thetacont2}), we have
\begin{align}
&\frac{|\mathbf{\Theta}+\mathbf{\Sigma}_{\mathbf{U}_1,\mathbf{U}_2}|}{|\mathbf{\Theta}|}\geq\frac{|\mathbf{\Sigma}_{\mathbf{S}_1,\mathbf{S}_2}+\mathbf{\Sigma}_{\mathbf{U}_1,\mathbf{U}_2}|}{|\mathbf{\Sigma}_{\mathbf{S}_1,\mathbf{S}_2}|},\label{eq:sub1}\\
&\frac{|\mathbf{\Theta}_2+\mathbf{\Sigma}_{\mathbf{U}_2}|}{|\mathbf{\Theta}+\mathbf{\Sigma}_{\mathbf{U}_1,\mathbf{U}_2}|}\geq\frac{1}{|\mathbf{\Theta}_1+\mathbf{\Sigma}_{\mathbf{U}_1}|}\geq\frac{1}{|\mathbf{D}_1+\mathbf{\Sigma}_{\mathbf{U}_1}|},\label{eq:sub2}
\end{align}
where (\ref{eq:sub1}) is due to the fact that $\frac{|\mathbf{A}_1+\mathbf{B}|}{|\mathbf{A}_1|}\geq\frac{|\mathbf{A}_2+\mathbf{B}|}{|\mathbf{A}_2|}$ for $\mathbf{A}_2\succeq\mathbf{A}_1\succ\mathbf{0}$ and $\mathbf{B}\succeq\mathbf{0}$, and the first inequality in (\ref{eq:sub2}) is a consequence of  Fischer's inequality.
Substituting (\ref{eq:sub1}) and (\ref{eq:sub2}) into (\ref{eq:identity}) yields
\begin{align*}
\frac{|\mathbf{\Sigma}_{\mathbf{S}_1,\mathbf{S}_2}||\mathbf{\Theta}_2+\mathbf{\Sigma}_{\mathbf{U}_2}|}{|\mathbf{\Theta}||\mathbf{D}_2+\mathbf{\Sigma}_{\mathbf{U}_2}|}\geq\frac{|\mathbf{\Sigma}_{\mathbf{S}_1,\mathbf{S}_2}+\mathbf{\Sigma}_{\mathbf{U}_1,\mathbf{U}_2}|}{|\mathbf{D}_1+\mathbf{\Sigma}_{\mathbf{U}_1}||\mathbf{D}_2+\mathbf{\Sigma}_{\mathbf{U}_2}|},
\end{align*}
from which Corollary \ref{cor:lowersimplified} follows immediately.
\end{IEEEproof}

It is also possible to derive this lower bound by taking a shortcut in the proof of Theorem \ref{thm:lower}.
\begin{IEEEproof}[Alternative Proof of Corollary \ref{cor:lowersimplified}]
Note that
\begin{align}
&I(\mathbf{V}^n_1;Y^n_1|\mathbf{V}^n_2)\nonumber\\
&=h(Y^n_1|\mathbf{V}^n_2)-h(Y^n_1|\mathbf{V}^n_1,\mathbf{V}^n_2)\nonumber\\
&\leq\frac{n}{2}\log(2\pi e(\alpha P+N_1))-h(Y^n_1|\mathbf{S}^n_1,\mathbf{S}^n_2)\label{eq:ineq1}\\
&=\frac{n}{2}\log(2\pi e(\alpha P+N_1))-h(Z^n_1)\nonumber\\
&=\frac{n}{2}\log\frac{\alpha P+N_1}{N_1}.\label{eq:tobecomb1}
\end{align}
On the other hand,
\begin{align}
&I(\mathbf{V}^n_1;Y^n_1|\mathbf{V}^n_2)\nonumber\\
&=h(\mathbf{V}^n_1|\mathbf{V}^n_2)-h(\mathbf{V}^n_1|\mathbf{V}^n_2,Y^n_1)\nonumber\\
&=\frac{n}{2}\log\frac{|2\pi e(\mathbf{\Sigma}_{\mathbf{S}_1,\mathbf{S}_2}+\mathbf{\Sigma}_{\mathbf{U}_1,\mathbf{U}_2})|}{|2\pi e(\mathbf{\Sigma}_{\mathbf{S}_2}+\mathbf{\Sigma}_{\mathbf{U}_2})|}-h(\mathbf{V}^n_1|\mathbf{V}^n_2,Y^n_1)\nonumber\\
&\geq\frac{n}{2}\log\frac{|2\pi e(\mathbf{\Sigma}_{\mathbf{S}_1,\mathbf{S}_2}+\mathbf{\Sigma}_{\mathbf{U}_1,\mathbf{U}_2})|}{|2\pi e(\mathbf{\Sigma}_{\mathbf{S}_2}+\mathbf{\Sigma}_{\mathbf{U}_2})|}-h(\mathbf{V}^n_1|Y^n_1)\label{eq:ineq2}\\
&\geq\frac{n}{2}\log\frac{|2\pi e(\mathbf{\Sigma}_{\mathbf{S}_1,\mathbf{S}_2}+\mathbf{\Sigma}_{\mathbf{U}_1,\mathbf{U}_2})|}{|2\pi e(\mathbf{\Sigma}_{\mathbf{S}_2}+\mathbf{\Sigma}_{\mathbf{U}_2})|}-h(\mathbf{V}^n_1|\hat{\mathbf{S}}^n_1)\nonumber\\
&\geq\frac{n}{2}\log\frac{|2\pi e(\mathbf{\Sigma}_{\mathbf{S}_1,\mathbf{S}_2}+\mathbf{\Sigma}_{\mathbf{U}_1,\mathbf{U}_2})|}{|2\pi e(\mathbf{\Sigma}_{\mathbf{S}_2}+\mathbf{\Sigma}_{\mathbf{U}_2})|}\nonumber\\
&\quad-\frac{n}{2}\log|2\pi e(\mathbf{\Theta}_1+\mathbf{\Sigma}_{\mathbf{U}_1})|\label{eq:invokelem1}\\
&\geq\frac{n}{2}\log\frac{|\mathbf{\Sigma}_{\mathbf{S}_1,\mathbf{S}_2}+\mathbf{\Sigma}_{\mathbf{U}_1,\mathbf{U}_2}|}{|\mathbf{\Sigma}_{\mathbf{S}_2}+\mathbf{\Sigma}_{\mathbf{U}_2}||\mathbf{D}_1+\mathbf{\Sigma}_{\mathbf{U}_1}|},\label{eq:tobecomb2}
\end{align}
where (\ref{eq:invokelem1}) follows from Lemma \ref{lem:entropybound}. Combining (\ref{eq:tobecomb1}) and (\ref{eq:tobecomb2}) yields
\begin{align*}
\frac{|\mathbf{\Sigma}_{\mathbf{S}_1,\mathbf{S}_2}+\mathbf{\Sigma}_{\mathbf{U}_1,\mathbf{U}_2}|}{|\mathbf{\Sigma}_{\mathbf{S}_2}+\mathbf{\Sigma}_{\mathbf{U}_2}||\mathbf{D}_1+\mathbf{\Sigma}_{\mathbf{U}_1}|}\leq\frac{\alpha P+N_1}{N_1},
\end{align*}
which, together with (\ref{eq:main1}), proves Corollary \ref{cor:lowersimplified}.
\end{IEEEproof}

In order for the inequalities in (\ref{eq:ineq1}) and (\ref{eq:ineq2}) to become equalities, we need to have
\begin{align}
&I(\mathbf{V}^n_1,\mathbf{V}^n_2;Y^n_1)=I(\mathbf{S}^n_1,\mathbf{S}^n_2;Y^n_1),\label{eq:para1}\\
&I(\mathbf{V}^n_1;\mathbf{V}^n_2|Y^n_1)=0.\label{eq:para2}
\end{align}
It will be seen that these two conditions provide important guidelines for constructing hybrid schemes that achieve the lower bound in Corollary \ref{cor:lowersimplified}.  Note that the derivation of this lower bound is based on a consideration of the scenario where $\mathbf{V}_2$ is provided to the strong receiver by a genie. Intuitively, a necessary condition for this lower bound to be tight is that the side information provided by the genie is superfluous, which is exactly the implication of (\ref{eq:para2}).

\section{the Vector-Scalar Case}\label{sec:upper}

We shall show in this section that the lower bound in Corollary \ref{cor:lowersimplified} is tight for the vector-scalar case, i.e., the scenario where the weak receiver wishes to reconstruct a  scalar source (i.e., $m_2=1$) under the mean squared error distortion constraint. In this special setup, we denote $\mathbf{S}_2,\mathbf{\Sigma}_{\mathbf{S}_2},\mathbf{D}_2,\mathbf{U}_2,\mathbf{\Sigma}_{\mathbf{U}_2}$ by $S_2,\sigma^2_{S_2},d_2,U_2,\sigma^2_{U_2}$, respectively.

\begin{theorem}\label{thm:vectorscalar}
\begin{align}
P(\mathbf{D}_1,d_2)&=\sup\limits_{\mathbf{\Sigma}_{\mathbf{U}_1,U_2}\succ \mathbf{0}}N_1\frac{|\mathbf{\Sigma}_{\mathbf{S}_1,S_2}+\mathbf{\Sigma}_{\mathbf{U}_1,U_2}|}{|\mathbf{D}_1+\mathbf{\Sigma}_{\mathbf{U}_1}|(d_2+\sigma^2_{U_2})}\nonumber\\
&\hspace{0.7in}+(N_2-N_1)\frac{\sigma^2_{S_2}+\sigma^2_{U_2}}{d_2+\sigma^2_{U_2}}-N_2.\label{eq:achievable}
\end{align}
\end{theorem}

\subsection{Upper Bound}\label{subsec:upper}

\begin{IEEEproof}[Proof of  Theorem \ref{thm:vectorscalar}]
To the end of proving Theorem \ref{thm:vectorscalar}, it suffices to show that the right-hand side of (\ref{eq:achievable}) is (asymptotically) achievable and consequently is an upper bound on  $P(\mathbf{D}_1,d_2)$. Our achievability argument is based on a hybrid scheme, which bears some resemblance to the one proposed by Puri \textit{et al.} in a different setting \cite{PRP02} (see also \cite{PPR08}). It will be seen that this hybrid scheme is semi-universal in the sense that the encoder only needs to know $N_1$ but not $N_2$. Let us first introduce a zero-mean random vector $\mathbf{S}_1(\gamma)$ and a zero-mean random variable $S_2(\gamma)$ that are jointly Gaussian. They are related with $(\mathbf{S}_1,S_2)$  via a backward Gaussian test channel $(\mathbf{S}_1,S_2)=(\mathbf{S}_1(\gamma)+\mathbf{Q}_1,S_2(\gamma)+Q_2)$, where $(\mathbf{Q}_1,Q_2)$ is independent of $(\mathbf{S}_1(\gamma),S_2(\gamma))$. The covariance matrix of  $(\mathbf{S}_1(\gamma),S_2(\gamma))$, parametrized by a scalar variable $\gamma$, is to be specified later. We assume that $(\mathbf{S}_1,S_2,\mathbf{S}_1(\gamma),S_2(\gamma))$ is independent of $(Z_1,Z_2)$. Note that we can write
\begin{align*}
\mathbf{S}_1(\gamma)&=\mathbb{E}[\mathbf{S}_1(\gamma)|\mathbf{S}_1,S_2,S_2(\gamma)]+\mathbf{W}_1\\
&=\mathbf{A}_1\mathbf{S}_1+\mathbf{a}_2S_2+\mathbf{a}_3S_2(\gamma)+\mathbf{W}_1,\\
S_2(\gamma)&=\mathbb{E}[S_2(\gamma)|\mathbf{S}_1,S_2]+W_2\\
&=\mathbf{b}^T_1\mathbf{S}_1+b_2S_2+W_2,
\end{align*}
where $\mathbf{W}_1$ is independent of $(\mathbf{S}_1,S_2,S_2(\gamma))$, and $W_2$ is independent of $(\mathbf{S}_1,S_2)$.
Next define
\begin{align*}
\tilde{\mathbf{S}}_1(\gamma)= \mathbf{A}_1\mathbf{S}_1+\mathbf{a}_2S_2+\mathbf{W}_1.
\end{align*}

\begin{figure*}[tb]
\begin{centering}
\includegraphics[width=11.5cm]{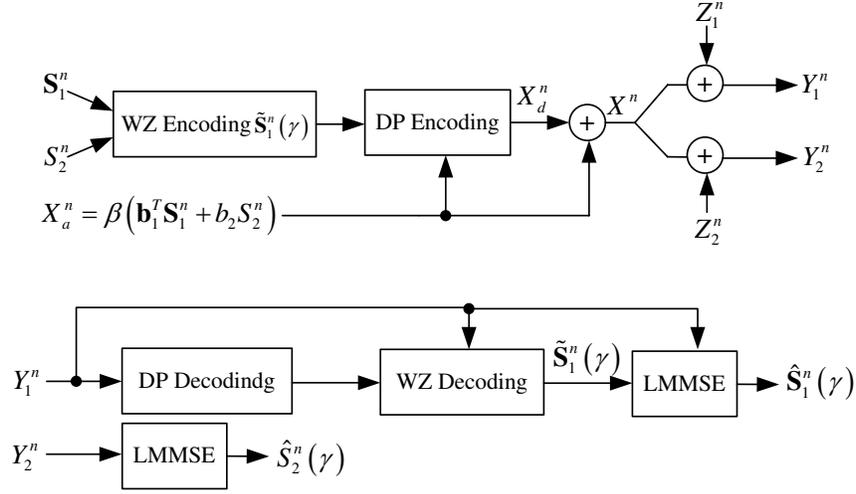}
\caption{An illustration of the hybrid scheme in Section \ref{subsec:upper}.\label{fig:hybrid1}}
\end{centering}
\end{figure*}

We are now in a position to describe the scheme (See Fig. \ref{fig:hybrid1}). Since the scheme is a combination of some well-known coding techniques, e.g., Wyner-Ziv codes  \cite{WynerZiv:76} and dirty paper codes \cite{Costa:83}, we only provide an outline of the encoding and decoding steps, and then focus on the condition that guarantees correct decoding.

\vspace{0.5cm}
\noindent\textbf{Encoding:} Let the channel input $X^n$, with average power $P(\gamma)$, be a superposition of an analog signal $X^n_a$ and a digital signal $X^n_d$ (i.e., $X^n=X^n_a+X^n_d$).  The analog portion is given by $X^n_a=\beta(\mathbf{b}^T_1\mathbf{S}^n_1+b_2S^n_2)$ for some non-negative number $\beta$ to be specified later. For the digital portion $X^n_d$, the encoder first uses a Wyner-Ziv code of rate $R$ with codewords generated according to $\tilde{\mathbf{S}}_1(\gamma)$, with $(\mathbf{S}^n_1,S^n_2)$ as the input, and with $Y^n_1\triangleq X^n_a+X^n_d+Z^n_1$ as the decoder side information; the encoder then determines the digital portion of the channel input $X^n_d$ to send the bin index of the chosen Wyner-Ziv codeword $\tilde{\mathbf{S}}^n_1(\gamma)$  by using a dirty paper code of rate $R$ with $X^n_a$ treated as the channel state information known at the encoder. We define $P_a=\mathbb{E}[(X_a)^2]$ and $P_d=\mathbb{E}[(X_d)^2]$, where $X_a\triangleq\beta(\mathbf{b}^T_1\mathbf{S}_1+b_2S_2)$ and $X_d$ are mutually independently zero-mean Gaussian random variables, and $P_a+P_d=P(\gamma)$.

\vspace{0.5cm}
\noindent\textbf{Decoding:} Receiver 1 first decodes the dirty paper code; it then further recovers $\tilde{\mathbf{S}}^n_1(\gamma)$ by decoding the Wyner-Ziv code with $Y^n_1$ as the side information. In view of the fact that the linear MMSE estimate of $\mathbf{S}_1$ based on $\tilde{\mathbf{S}}_1(\gamma)$ and $Y_1\triangleq X_a+X_d+Z_1$ is $\hat{\mathbf{S}}_1(\gamma)\triangleq\tilde{\mathbf{S}}_1(\gamma)+\beta^{-1}\mathbf{a}_3Y_1$, Receiver 1 can use $\hat{\mathbf{S}}^n_1(\gamma)\triangleq\tilde{\mathbf{S}}^n_1(\gamma)+\beta^{-1}\mathbf{a}_3Y^n_1$ as the reconstruction of $\mathbf{S}^n_1$.
Since the linear MMSE estimate of $S_2$ based on $Y_2\triangleq X_a+X_d+Z_2$ is $\hat{S}_2(\gamma)\triangleq\rho Y_2$ with $\rho=\mathbb{E}[S_2X_a](P(\gamma)+N_2)^{-1}$, Receiver 2 can simply use $\hat{S}^n_2(\gamma)\triangleq\rho Y^n_2$ as the reconstruction of $S^n_2$, where $Y^n_2= X^n_a+X^n_d+Z^n_2$; the resulting distortion is denoted by $d_2(\gamma)$.

\vspace{0.5cm}
\noindent\textbf{Coding Parameters:}
For a given covariance matrix of $(\mathbf{S}_1(\gamma),S_2(\gamma))$, three parameters $\beta$, $P_d$, and $R$ still need to be specified for the aforedescribed scheme. Equivalently, we shall specify $\beta$, $P(\gamma)$, and $R$, since $\beta$ determines $P_a$ and $P_d=P(\gamma)-P_a$. Let us first choose $P(\gamma)$ such that
\begin{align}
I(\mathbf{S}_1,S_2;\mathbf{S}_1(\gamma),S_2(\gamma))=\frac{1}{2}\log\frac{P(\gamma)+N_1}{N_1}.\label{eqn:Information}
\end{align}
The parameter $\beta$ is then chosen such that
\begin{align}
I(X_a;Y_1)=I(\mathbf{S}_1,S_2;S_2(\gamma)),\label{eq:statisticequi}
\end{align}
which is always possible because
\begin{align*}
I(\mathbf{S}_1,S_2;S_2(\gamma))&\leq I(\mathbf{S}_1,S_2;\mathbf{S}_1(\gamma),S_2(\gamma))\\
&=\frac{1}{2}\log\frac{P(\gamma)+N_1}{N_1},
\end{align*}
and one can let $I(X_a;Y_1)$ take any value in $[0,\frac{1}{2}\log\frac{P(\gamma)+N_1}{N_1}]$ by varying $\beta$.
Finally set
\begin{align}
R=I(\mathbf{S}_1,S_2;\tilde{\mathbf{S}}_1(\gamma)|Y_1).\label{eq:WZrate}
\end{align}
Now the scheme is fully specified for any given covariance matrix of $(\mathbf{S}_1(\gamma),S_2(\gamma))$.

\vspace{0.5cm}
\noindent\textbf{Conditions for Correct Decoding:} The Wyner-Ziv code and the dirty paper code need to be decoded correctly at Receiver 1. It is easily seen that the Wyner-Ziv code is ensured to be decoded correctly by (\ref{eq:WZrate}), and thus we focus on the decodability of dirty paper code. First note that (\ref{eq:statisticequi}), together with the fact that $I(X_a;Y_1)=I(\mathbf{S}_1,S_2;Y_1)$, implies that $I(\mathbf{S}_1,S_2;Y_1)=I(\mathbf{S}_1,S_2;S_2(\gamma))$; moreover, since both $X_d+Z_1$ and $W_2$, which are Gaussian random variables, are independent of $(\mathbf{S}_1,S_2)$, it follows that the joint distributions of $(\mathbf{S}_1,S_2,\beta^{-1}Y_1)$ and $(\mathbf{S}_1,S_2,S_2(\gamma))$ are identical, which, in view of the fact that $\mathbf{W}_1$ is independent of $(\mathbf{S}_1,S_2,S_2(\gamma),Y_1)$, further implies that the joint distributions of $(\mathbf{S}_1,S_2,\tilde{\mathbf{S}}_1(\gamma),\beta^{-1}Y_1)$ and $(\mathbf{S}_1,S_2,\tilde{\mathbf{S}}_1(\gamma),S_2(\gamma))$ are identical\footnote{We have implicitly assumed that $\mathbb{E}[(\mathbf{b}^T_1\mathbf{S}_1+b_2S_2)^2]>0$ (which implies that the $P_a$ and the $\beta$ determined by (\ref{eq:statisticequi}) are positive). For the degenerate case $\mathbf{b}^T_1\mathbf{S}_1+b_2S_2=0$ (which is possible if and only if $S_2(\gamma)=0$), one can simply set $X_a=0$ and $\beta^{-1}Y_1=0$.
}. Therefore, we have
\begin{align}
R=I(\mathbf{S}_1,S_2;\tilde{\mathbf{S}}_1(\gamma)|S_2(\gamma)).\label{eq:DPrate}
\end{align}
Furthermore, note that
\begin{align*}
&I(\mathbf{S}_1,S_2;\tilde{\mathbf{S}}_1(\gamma)|S_2(\gamma))\nonumber\\
&=I(\mathbf{S}_1,S_2;\tilde{\mathbf{S}}_1(\gamma),S_2(\gamma))-I(\mathbf{S}_1,S_2;S_2(\gamma))\nonumber\\
&=I(\mathbf{S}_1,S_2;\mathbf{S}_1(\gamma),S_2(\gamma))-I(X_a;Y_1)\nonumber\\
&=\frac{1}{2}\log\frac{P(\gamma)+N_1}{N_1}-\frac{1}{2}\log\frac{P(\gamma)+N_1}{P_d+N_1}\nonumber\\
&=\frac{1}{2}\log\frac{P_d+N_1}{N_1},\nonumber
\end{align*}
which, together with (\ref{eq:DPrate}), ensures that Receiver 1 can correctly decode the dirty paper code.

\vspace{0.5cm}
\noindent\textbf{Optimizing the Covariance Matrix of $(\mathbf{S}_1(\gamma),S_2(\gamma))$:} Now only the covariance matrix of $(\mathbf{S}_1(\gamma),S_2(\gamma))$ remains to be specified. To this end we formulate the following maximization problem. It will become clear that this maximization problem is motivated by the lower bound in Corollary \ref{cor:lowersimplified}. In particular, it will be seen that the hybrid scheme and the remote sources induced by the optimal solution (and the associated Lagrangian multipliers) of this maximization problem possess the desired properties (see (\ref{eq:para1}) and (\ref{eq:para2})).

Given $\gamma\in(0,\infty)$, let $\mathbf{\Theta}(\gamma)$ denote the solution\footnote{Note that $\mathbf{\Theta}(\gamma)$ must be positive definite. Since $\log|\cdot|$ is strictly concave over the domain of positive definite matrices, it follows that $\mathbf{\Theta}(\gamma)$ is uniquely defined.} to
\begin{align}
\max\limits_{\mathbf{\Theta}}&\log|\mathbf{\Theta}|\label{eq:opt}\\
\mbox{subject to}\quad&\mathbf{\Theta}_{1}\preceq\mathbf{D}_1,\nonumber\\
&\theta_{2}\leq\gamma,\nonumber\\
&\mathbf{0}\preceq\mathbf{\Theta}\preceq\mathbf{\Sigma}_{\mathbf{S}_1,S_2},\nonumber
\end{align}
where $\mathbf{\Theta}_{1}$ is the first $m_1\times m_1$ diagonal submatrix of $\mathbf{\Theta}$, and $\theta_{2}$ is the $(m_1+1,m_1+1)$ entry of $\mathbf{\Theta}$.  It can be shown (see Appendix \ref{app:theta}) that $\mathbf{\Theta}(\gamma)$ is a continuous function of $\gamma$.
We denote the first $m_1\times m_1$ diagonal submatrix of $\mathbf{\Theta}(\gamma)$ by $\mathbf{\Theta}_{1}(\gamma)$, and the $(m_1+1,m_1+1)$ entry of $\mathbf{\Theta}(\gamma)$ by $\theta_{2}(\gamma)$.
Now choose the covariance matrix of $(\mathbf{S}_1(\gamma),S_2(\gamma))$ to be $\mathbf{\Sigma}_{\mathbf{S}_1,S_2}-\mathbf{\Theta}(\gamma)$; as a consequence, the covariance matrix of $\mathbf{S}_1(\gamma)$ is $\mathbf{\Sigma}_{\mathbf{S}_1}-\mathbf{\Theta}_{1}(\gamma)$, and the variance of $S_2(\gamma)$ is $\sigma^2_{S_2}-\theta_{2}(\gamma)$.
Accordingly, (\ref{eqn:Information}) reduces to
\begin{align}
\frac{1}{2}\log\frac{|\mathbf{\Sigma}_{\mathbf{S}_1,S_2}|}{|\mathbf{\Theta}(\gamma)|}=\frac{1}{2}\log\frac{P(\gamma)+N_1}{N_1}.\label{eq:direct}
\end{align}

\vspace{0.5cm}
\noindent\textbf{Evaluating the Distortions and the Transmit Power:} For the distortion at Receiver 1, it is readily seen that
\begin{align}
&\mathbb{E}[(\mathbf{S}_1-\hat{\mathbf{S}}_1(\gamma))(\mathbf{S}_1-\hat{\mathbf{S}}_1(\gamma))^T]\nonumber\\
&=\mathbb{E}[(\mathbf{S}_1-\mathbf{S}_1(\gamma))(\mathbf{S}_1-\mathbf{S}_1(\gamma))^T]\label{eqn:equivalenceS1}\\
&=\mathbf{\Theta}_{1}(\gamma)\nonumber\\
&\preceq\mathbf{D}_1,\nonumber
\end{align}
where (\ref{eqn:equivalenceS1}) is true because the joint distributions of $(\mathbf{S}_1,\hat{\mathbf{S}}_1(\gamma))$ and $(\mathbf{S}_1,\mathbf{S}_1(\gamma))$ are identical (which is further due to the fact that the joint distributions of $(\mathbf{S}_1,\tilde{\mathbf{S}}_1(\gamma),\beta^{-1}Y_1)$ and $(\mathbf{S}_1,\tilde{\mathbf{S}}_1(\gamma),S_2(\gamma))$ are identical).
It is worth noting that the linear MMSE estimate of $(\mathbf{S}_1,S_2)$ based on $(\tilde{\mathbf{S}}_1(\gamma),Y_1)$ is $(\hat{\mathbf{S}}_1(\gamma),\beta^{-1}Y_1)$. In view of this fact, Receiver 1 can use $(\hat{\mathbf{S}}^n_1(\gamma),\beta^{-1}Y^n_1)$ as the reconstruction of $(\mathbf{S}^n_1,S^n_2)$. Since the joint distributions of $(\mathbf{S}_1,S_2,\hat{\mathbf{S}}_1(\gamma),\beta^{-1}Y_1)$ and $(\mathbf{S}_1,S_2,\mathbf{S}_1(\gamma),S_2(\gamma))$ are identical, we have
\begin{align}
&\mathbb{E}[(\mathbf{S}^T_1-\hat{\mathbf{S}}^T_1(\gamma),S^T_2-\beta^{-1}Y^T_1)^T\nonumber\\
&\quad(\mathbf{S}^T_1-\hat{\mathbf{S}}^T_1(\gamma),S^T_2-\beta^{-1}Y^T_1)]\nonumber\\
&=\mathbb{E}[(\mathbf{S}^T_1-\hat{\mathbf{S}}^T_1(\gamma),S^T_2-S^T_2(\gamma))^T\nonumber\\
&\hspace{0.32in}(\mathbf{S}^T_1-\hat{\mathbf{S}}^T_1(\gamma),S^T_2-S^T_2(\gamma))]\nonumber\\
&=\mathbf{\Theta}(\gamma).\label{eq:d2p}
\end{align}
Therefore, $\gamma$ can be interpreted as an auxiliary constraint on the reconstruction distortion for $S^n_2$ at Receiver 1, and $\mathbf{\Theta}(\gamma)$ is the actual covariance distortion achieved at Receiver 1 for reconstructing $(\mathbf{S}^n_1,S^n_2)$.

Note that $P(\gamma)$ is a continuous function of $\mathbf{\Theta}(\gamma)$ (which is implied by (\ref{eq:direct})) and consequently is a continuous function of $\gamma$ for $\gamma\in(0,\infty)$. Moreover, it can be verified that
\begin{align}
&\frac{1}{2}\log\frac{\sigma^2_{S_2}}{d_2(\gamma)}\nonumber\\
&=I(S_2;Y_2)\nonumber\\
&=\frac{1}{2}\log\frac{P(\gamma)+N_2}{\mathbb{E}[(Y_2-\mathbb{E}[Y_2|S_2])^2]}\nonumber\\
&=\frac{1}{2}\log\frac{P(\gamma)+N_2}{\mathbb{E}[(Y_1-\mathbb{E}[Y_1|S_2])^2]+N_2-N_1}\nonumber\\
&=\frac{1}{2}\log(P(\gamma)+N_2)-\frac{1}{2}\log\left(\frac{1}{2\pi e}e^{2h(Y_1|S_2)}+N_2-N_1\right)\nonumber\\
&=\frac{1}{2}\log(P(\gamma)+N_2)\nonumber\\
&\quad-\frac{1}{2}\log\left(\frac{1}{2\pi e}e^{2(h(Y_1)-I(S_2;Y_1))}+N_2-N_1\right)\nonumber\\
&=\frac{1}{2}\log(P(\gamma)+N_2)\nonumber\\
&\quad-\frac{1}{2}\log\left(\frac{P(\gamma)+N_1}{2\pi e\sigma^2_{S_2}}e^{2h(S_2|Y_1)}+N_2-N_1\right)\nonumber\\
&=\frac{1}{2}\log(P(\gamma)+N_2)\nonumber\\
&\quad-\frac{1}{2}\log\left(\frac{(P(\gamma)+N_1)\theta_2(\gamma)}{\sigma^2_{S_2}}+N_2-N_1\right),\label{eq:d2pderive}
\end{align}
where (\ref{eq:d2pderive}) is due to the fact that $h(S_2|Y_1)=\frac{1}{2}\log(2\pi e\theta_2(\gamma))$ (which is implied by (\ref{eq:d2p})). Hence,
\begin{align*}
d_2(\gamma)=\frac{(P(\gamma)+N_1)\theta_2(\gamma)+(N_2-N_1)\sigma^2_{S_2}}{P(\gamma)+N_2}.
\end{align*}
Note that both $P(\gamma)$ and $\theta_2(\gamma)$ are continuous in $\gamma$; furthermore, $P(\gamma)$ and $\theta_2(\gamma)$ tend to infinity and zero, respectively, as $\gamma\rightarrow0$. Therefore, $d_2(\gamma)$ is a continuous function of $\gamma$ for $\gamma\in(0,\infty)$, and $d_2(\gamma)$ tends to zero as
$\gamma\rightarrow 0$.

We shall show that
\begin{align}
P(\gamma)&\leq\sup\limits_{\mathbf{\Sigma}_{\mathbf{U}_1,U_2}\succ \mathbf{0}}N_1\frac{|\mathbf{\Sigma}_{\mathbf{S}_1,S_2}+\mathbf{\Sigma}_{\mathbf{U}_1,U_2}|}{|\mathbf{D}_1+\mathbf{\Sigma}_{\mathbf{U}_1}|(d_2(\gamma)+\sigma^2_{U_2})}\nonumber\\
&\hspace{0.7in}+(N_2-N_1)\frac{\sigma^2_{S_2}+\sigma^2_{U_2}}{d_2(\gamma)+\sigma^2_{U_2}}-N_2 \label{eq:tobeproved}
\end{align}
for $\gamma\in(0,\infty)$.
To this end we revisit the maximization problem in (\ref{eq:opt}). Note that $\mathbf{\Theta}(\gamma)$ must satisfy the following KKT conditions \cite{Bertsekas99}
\begin{align}
&\mathbf{\Theta}^{-1}(\gamma)-\mathbf{\Lambda}-\mathbf{M}=\mathbf{0},\label{eq:KKT1}\\
&\mathbf{\Lambda}_1(\mathbf{D}_1-\mathbf{\Theta}_{1}(\gamma))=\mathbf{0},\nonumber\\
&\lambda_2(\gamma-\theta_{2}(\gamma))=0,\nonumber\\
&\mathbf{M}(\mathbf{\Sigma}_{\mathbf{S}_1,S_2}-\mathbf{\Theta}(\gamma))=\mathbf{0},\label{eq:KKT3}
\end{align}
where $\mathbf{M}\succeq\mathbf{0}$, $\mathbf{\Lambda}_1\succeq\mathbf{0}$, $\lambda_2\geq 0$, and $\mathbf{\Lambda}=\mbox{diag}(\mathbf{\Lambda}_1,\lambda_2)$.
Let $\mathbf{\Xi}_1\mathbf{\Pi}_1\mathbf{\Xi}^T_1$ be the eigenvalue decomposition of $\mathbf{\Lambda}_1$, where $\mathbf{\Xi}_1$ is a unitary matrix, and $\mathbf{\Pi}_1=\mbox{diag}(\pi_1,\cdots,\pi_r,0,\cdots,0)$ with $\pi_i>0$, $i=1,\cdots,r$. Define $\mathbf{\Xi}=\mbox{diag}(\mathbf{\Xi}_1,1)$ and $\mathbf{\Pi}=\mbox{diag}(\mathbf{\Pi}_1,\lambda_2)$. Let $\mathbf{\Pi}'_{\epsilon}$ be a positive semidefinite diagonal matrix obtained by subtracting $\epsilon$ from each positive diagonal entry of $\mathbf{\Pi}$, where $\epsilon$ is an arbitrary positive number smaller than the minimum non-zero diagonal entry of $\mathbf{\Pi}$.
Since $\mathbf{\Theta}\succ\mathbf{0}$, it follows that $\mathbf{\Xi}^T\mathbf{\Theta}^{-1}(\gamma)\mathbf{\Xi}$ is positive definite. Moreover, in view of (\ref{eq:KKT1}), we have $\mathbf{\Xi}^T\mathbf{\Theta}^{-1}(\gamma)\mathbf{\Xi}-\mathbf{\Pi}=\mathbf{\Xi}^T\mathbf{M}\mathbf{\Xi}\succeq\mathbf{0}$.
Therefore, $\mathbf{\Xi}^T\mathbf{\Theta}^{-1}(\gamma)\mathbf{\Xi}-\mathbf{\Pi}'_{\epsilon}$ is positive definite when $\epsilon$ is sufficiently small.
For any $\epsilon$ with $\mathbf{\Xi}^T\mathbf{\Theta}^{-1}(\gamma)\mathbf{\Xi}-\mathbf{\Pi}'_{\epsilon}\succ\mathbf{0}$, we choose a positive number $\epsilon'$, which is a function of $\epsilon$ and tends to zero as $\epsilon\rightarrow 0$, such that
\begin{align*}
\mathbf{\Xi}^T\mathbf{\Theta}^{-1}(\gamma)\mathbf{\Xi}-\mathbf{\Pi}_{\epsilon}\succ\mathbf{0},
\end{align*}
where $\mathbf{\Pi}_{\epsilon}$ is a positive definite diagonal matrix obtained by adding $\epsilon'$ to each zero diagonal entry of $\mathbf{\Pi}'_{\epsilon}$. Now let $\mathbf{\Lambda}_{\epsilon}=\mathbf{\Xi}\mathbf{\Pi}_{\epsilon}\mathbf{\Xi}^T$ and $\mathbf{\Sigma}_{\mathbf{U}_{1,\epsilon},U_{2,\epsilon}}=\mathbf{\Lambda}^{-1}_{\epsilon}-\mathbf{\Theta}(\gamma)$. Note that
\begin{align*}
&\mathbf{\Xi}^T\mathbf{\Theta}^{-1}(\gamma)\mathbf{\Xi}-\mathbf{\Pi}_{\epsilon}\succ\mathbf{0}\\
&\Rightarrow\mathbf{\Theta}^{-1}(\gamma)\succ\mathbf{\Lambda}_{\epsilon}\\
&\Rightarrow\mathbf{\Lambda}^{-1}_{\epsilon}\succ\mathbf{\Theta}(\gamma).
\end{align*}
Therefore, $\mathbf{\Sigma}_{\mathbf{U}_{1,\epsilon},U_{2,\epsilon}}$ is positive definite when $\epsilon$ is sufficiently small.

Let $\mathbf{U}_{1,\epsilon}$ and $U_{2,\epsilon}$ be jointly Gaussian with mean zero and covariance matrix $\mathbf{\Sigma}_{\mathbf{U}_{1,\epsilon},U_{2,\epsilon}}$, where $\mathbf{U}_{1,\epsilon}$ is an $m_1\times 1$ Gaussian random vector with covariance matrix $\mathbf{\Sigma}_{\mathbf{U}_{1,\epsilon}}$ (which is the first $m_1\times m_1$ diagonal submatrix of $\mathbf{\Sigma}_{\mathbf{U}_{1,\epsilon},U_{2,\epsilon}}$) and $U_{2,\epsilon}$ is a Gaussian random variable with variance $\sigma^2_{U_{2,\epsilon}}$ (which is the $(m_1+1,m_1+1)$ entry of $\mathbf{\Sigma}_{\mathbf{U}_{1,\epsilon},U_{2,\epsilon}}$). We assume that $(\mathbf{U}_{1,\epsilon}, U_{2,\epsilon})$ is independent of $(\mathbf{S}_1,S_2,\mathbf{S}_1(\gamma),S_2(\gamma),Z_1,Z_2)$.

Note that
\begin{align}
&\lim\limits_{\epsilon\rightarrow 0}\frac{|\mathbf{\Sigma}_{\mathbf{S}_1,S_2}+\mathbf{\Sigma}_{\mathbf{U}_{1,\epsilon},U_{2,\epsilon}}|}{|\mathbf{\Theta}(\gamma)+\mathbf{\Sigma}_{\mathbf{U}_{1,\epsilon},U_{2,\epsilon}}|}\nonumber\\
&=\lim\limits_{\epsilon\rightarrow 0}\frac{|\mathbf{\Sigma}_{\mathbf{S}_1,S_2}+\mathbf{\Lambda}^{-1}_{\epsilon}-\mathbf{\Theta}(\gamma)|}{|\mathbf{\Lambda}^{-1}_{\epsilon}|}\nonumber\\
&=\lim\limits_{\epsilon\rightarrow 0}|\mathbf{\Lambda}_{\epsilon}\mathbf{\Sigma}_{\mathbf{S}_1,S_2}+\mathbf{I}-\mathbf{\Lambda}_{\epsilon}\mathbf{\Theta}(\gamma)|\nonumber\\
&=|\mathbf{\Lambda}\mathbf{\Sigma}_{\mathbf{S}_1,S_2}+\mathbf{I}-\mathbf{\Lambda}\mathbf{\Theta}(\gamma)|\nonumber\\
&=|\mathbf{\Theta}^{-1}(\gamma)\mathbf{\Sigma}_{\mathbf{S}_1,S_2}-\mathbf{M}\mathbf{\Sigma}_{\mathbf{S}_1,S_2}+\mathbf{M}\mathbf{\Theta}(\gamma)|\label{eq:useKKT1}\\
&=\frac{|\mathbf{\Sigma}_{\mathbf{S},S_2}|}{|\mathbf{\Theta}(\gamma)|}\label{eq:useKKT3}\\
&=\frac{P(\gamma)+N_1}{N_1},\label{eq:tobeused1}
\end{align}
where (\ref{eq:useKKT1}) and (\ref{eq:useKKT3}) are due to (\ref{eq:KKT1}) and (\ref{eq:KKT3}), respectively. Moreover, by the definition of $\mathbf{\Sigma}_{\mathbf{U}_{1,\epsilon},U_{2,\epsilon}}$, we have
\begin{align}
&\mathbf{\Theta}(\gamma)+\mathbf{\Sigma}_{\mathbf{U}_{1,\epsilon},U_{2,\epsilon}}\nonumber\\
&=\mbox{diag}(\mathbf{\Theta}_{1}(\gamma)+\mathbf{\Sigma}_{\mathbf{U}_{1,\epsilon}}, \theta_{2}(\gamma)+\sigma^2_{U_{2,\epsilon}}).\label{eq:diag}
\end{align}

It is clear that
\begin{align*}
I(S_2+U_{2,\epsilon};Y_2)&=h(Y_2)-h(Y_2|S_2+U_{2,\epsilon})\\
&=\frac{1}{2}\log\frac{P(\gamma)+N_2}{\alpha_{\epsilon}P(\gamma)+N_2},
\end{align*}
where $\alpha_{\epsilon}=\frac{1}{P(\gamma)}\mathbb{E}[(X-\mathbb{E}[X|S_2+U_{2,\epsilon}])^2]$.
On the other hand,
\begin{align*}
I(S_2+U_{2,\epsilon};Y_2)&=h(S_2+U_{2,\epsilon})-h(S_2+U_{2,\epsilon}|Y_2)\\
&=\frac{1}{2}\log\frac{\sigma^2_{S_2}+\sigma^2_{U_{2,\epsilon}}}{d_2(\gamma)+\sigma^2_{U_{2,\epsilon}}}.
\end{align*}
Therefore,
\begin{align}
\frac{\sigma^2_{S_2}+\sigma^2_{U_{2,\epsilon}}}{d_2(\gamma)+\sigma^2_{U_{2,\epsilon}}}=\frac{P(\gamma)+N_2}{\alpha_{\epsilon}P(\gamma)+N_2}.\label{eq:combine1}
\end{align}
Note that
\begin{align}
&I(\mathbf{S}_1+\mathbf{U}_{1,\epsilon};\mathbf{S}_1(\gamma),S_2(\gamma)|S_2+U_{2,\epsilon})\nonumber\\
&=I(\mathbf{S}_1+\mathbf{U}_{1,\epsilon}, S_2+U_{2,\epsilon};\mathbf{S}_1(\gamma),S_2(\gamma))\nonumber\\
&\quad-I(S_2+U_{2,\epsilon};\mathbf{S}_1(\gamma),S_2(\gamma))\nonumber\\
&=I(\mathbf{S}_1+\mathbf{U}_{1,\epsilon}, S_2+U_{2,\epsilon};\mathbf{S}_1(\gamma),S_2(\gamma))\nonumber\\
&\quad-I(S_2+U_{2,\epsilon};S_2(\gamma))\label{eq:condind2}\\
&=I(\mathbf{S}_1+\mathbf{U}_{1,\epsilon}, S_2+U_{2,\epsilon};\mathbf{S}_1(\gamma),S_2(\gamma))-I(S_2+U_{2,\epsilon};Y_1)\nonumber\\
&=\frac{1}{2}\log\frac{|\mathbf{\Sigma}_{\mathbf{S}_1,S_2}+\mathbf{\Sigma}_{\mathbf{U}_{1,\epsilon},U_{2,\epsilon}}|}{|\mathbf{\Theta}(\gamma)+\mathbf{\Sigma}_{\mathbf{U}_{1,\epsilon},U_{2,\epsilon}}|}\nonumber\\
&\quad-\frac{1}{2}\log\frac{P(\gamma)+N_1}{\alpha_{\epsilon} P(\gamma)+N_1},\label{eq:tbj3}
\end{align}
where  (\ref{eq:condind2}) is due to (\ref{eq:diag}).
On the other hand,
\begin{align}
&I(\mathbf{S}_1+\mathbf{U}_{1,\epsilon};\mathbf{S}_1(\gamma),S_2(\gamma)|S_2+U_{2,\epsilon})\nonumber\\
&=h(\mathbf{S}_1+\mathbf{U}_{1,\epsilon}|S_2+U_{2,\epsilon})\nonumber\\
&\quad-h(\mathbf{S}+\mathbf{U}_{1,\epsilon}|S_2+U_{2,\epsilon},\mathbf{S}_1(\gamma),S_2(\gamma))\nonumber\\
&=h(\mathbf{S}_1+\mathbf{U}_{1,\epsilon}|S_2+U_{2,\epsilon})-h(\mathbf{S}+\mathbf{U}_{1,\epsilon}|\mathbf{S}_1(\gamma))\label{eq:condind}\\
&=\frac{1}{2}\log\frac{|\mathbf{\Sigma}_{\mathbf{S}_1,S_2}+\mathbf{\Sigma}_{\mathbf{U}_{1,\epsilon},U_{2,\epsilon}}|}{|\mathbf{\Theta}_{1}(\gamma)+\mathbf{\Sigma}_{\mathbf{U}_{1,\epsilon}}|(\sigma^2_{S_2}+\sigma^2_{U_{2,\epsilon}})},\label{eq:combine2}
\end{align}
where (\ref{eq:condind}) is due to (\ref{eq:diag}).
 Combining (\ref{eq:combine2}) and (\ref{eq:tbj3}) gives
\begin{align*}
&\frac{|\mathbf{\Sigma}_{\mathbf{S}_1,S_2}+\mathbf{\Sigma}_{\mathbf{U}_{1,\epsilon},U_{2,\epsilon}}|}{|\mathbf{\Theta}_{1}(\gamma)+\mathbf{\Sigma}_{\mathbf{U}_{1,\epsilon}}|(\sigma^2_{S_2}+\sigma^2_{U_{2,\epsilon}})}\\
&=\frac{|\mathbf{\Sigma}_{\mathbf{S}_1,S_2}+\mathbf{\Sigma}_{\mathbf{U}_{1,\epsilon},U_{2,\epsilon}}|(\alpha_{\epsilon} P(\gamma)+N_1)}{|\mathbf{\Theta}(\gamma)+\mathbf{\Sigma}_{\mathbf{U}_{1,\epsilon},U_{2,\epsilon}}|(P(\gamma)+N_1)},
\end{align*}
which, together with (\ref{eq:tobeused1}) and (\ref{eq:combine1}), implies that
\begin{align}
P(\gamma)&=\lim\limits_{\epsilon\rightarrow 0}N_1\frac{|\mathbf{\Sigma}_{\mathbf{S}_1,S_2}+\mathbf{\Sigma}_{\mathbf{U}_{1,\epsilon},U_{2,\epsilon}}|}{|\mathbf{\Theta}_{1}(\gamma)+\mathbf{\Sigma}_{\mathbf{U}_{1,\epsilon}}||d_2(\gamma)+\sigma^2_{U_{2,\epsilon}}|}\nonumber\\
&\hspace{0.37in}+(N_2-N_1)\frac{\sigma^2_{S_2}+\sigma^2_{U_{2,\epsilon}}}{d_2(\gamma)+\sigma^2_{U_{2,\epsilon}}}-N_2.\label{eq:Pxi}
\end{align}
Note that
\begin{align*}
&\mathbf{\Lambda}_1(\mathbf{D}_1-\mathbf{\Theta}_{1}(\gamma))=\mathbf{0}\\
&\Rightarrow\mathbf{\Xi}^T_1\mathbf{\Lambda}_1(\mathbf{D}_1-\mathbf{\Theta}_{1}(\gamma))\mathbf{\Xi}_1=\mathbf{0}\\
&\Rightarrow\mathbf{\Pi}_1\mathbf{\Xi}^T_1(\mathbf{D}_1-\mathbf{\Theta}_{1}(\gamma))\mathbf{\Xi}_1=\mathbf{0},
\end{align*}
which further implies that $\mathbf{\Xi}^T_1(\mathbf{D}_1-\mathbf{\Theta}_{1}(\gamma))\mathbf{\Xi}_1$ is of the form $\mbox{diag}(\mathbf{0}_{r\times r},\mathbf{A})$,
where $\mathbf{0}_{r\times r}$ denotes an $r\times r$ all-zero matrix. Also note that $\mathbf{\Xi}^T_1\mathbf{\Sigma}_{\mathbf{U}_{1,\epsilon}}\mathbf{\Xi}_1=\mathbf{\Pi}^{-1}_{1,\epsilon}-\mathbf{\Xi}^T_1\mathbf{\Theta}_{1}(\gamma)\mathbf{\Xi}_1$. Therefore,
\begin{align}
&\lim\limits_{\epsilon\rightarrow 0}\frac{|\mathbf{\Theta}_{1}(\gamma)+\mathbf{\Sigma}_{\mathbf{U}_{1,\epsilon}}|}{|\mathbf{D}_{1}+\mathbf{\Sigma}_{\mathbf{U}_{1,\epsilon}}|}\nonumber\\
&=\lim\limits_{\epsilon\rightarrow 0}\frac{|\mathbf{\Xi}^T_1\mathbf{\Theta}_{1}(\gamma)\mathbf{\Xi}_1+\mathbf{\Xi}^T_1\mathbf{\Sigma}_{\mathbf{U}_{1,\epsilon}}\mathbf{\Xi}_1|}{|\mathbf{\Xi}^T_1\mathbf{D}_{1}\mathbf{\Xi}_1+\mathbf{\Xi}^T_1\mathbf{\Sigma}_{\mathbf{U}_{1,\epsilon}}\mathbf{\Xi}_1|}\nonumber\\
&=\lim\limits_{\epsilon\rightarrow 0}\frac{|\mathbf{\Pi}^{-1}_{1,\epsilon}|}{|\mathbf{\Pi}^{-1}_{1,\epsilon}+\mathbf{\Xi}^T_1(\mathbf{D}_{1}-\mathbf{\Theta}_{1}(\gamma))\mathbf{\Xi}_1|}\nonumber\\
&=\lim\limits_{\epsilon\rightarrow 0}\frac{|\mathbf{\Pi}^{-1}_{1,\epsilon}|}{|\mathbf{\Pi}^{-1}_{1,\epsilon}+\mbox{diag}(\mathbf{0}_{r\times r},\mathbf{A})|}\nonumber\\
&=1.\label{eq:one}
\end{align}
Now one can readily prove (\ref{eq:tobeproved}) by combining (\ref{eq:Pxi}) and  (\ref{eq:one}). This completes the proof of Theorem \ref{thm:vectorscalar} for the case $d_2\leq d_2(\sigma^2_{S_2})$.

By restricting $\mathbf{\Sigma}_{\mathbf{U}_1,U_2}$ to the form $\mbox{diag}(\mathbf{\Sigma}_{\mathbf{U}_1},\sigma^2_{U_2})$ and letting  $\sigma^2_{U_2}\rightarrow\infty$, we can obtain the following lower bound from Corollary \ref{cor:lowersimplified}:
\begin{align}
P(\mathbf{D}_1,d_2)\geq \sup\limits_{\mathbf{\Sigma}_{\mathbf{U}_1}\succ0}N_1\frac{|\mathbf{\Sigma}_{\mathbf{S}_1}+\mathbf{\Sigma}_{\mathbf{U}_1}|}{|\mathbf{D}_1+\mathbf{\Sigma}_{\mathbf{U}_1}|}-N_1.\label{eq:separation}
\end{align}
Note that if $\gamma>\sigma^2_{S_2}$, then $\mathbf{\Theta}(\gamma)=\mathbf{\Theta}(\sigma^2_{S_2})$ and $d_2(\gamma)=d_2(\sigma^2_{S_2})$; moreover, in this case we have $\lambda_2=0$ (which implies that $\sigma^2_{U_{2,\epsilon}}$ tends to infinity as $\epsilon\rightarrow 0$), and consequently
\begin{align*}
P(\gamma)&=\lim\limits_{\epsilon\rightarrow 0}N_1\frac{|\mathbf{\Sigma}_{\mathbf{S}_1,S_2}+\mathbf{\Sigma}_{\mathbf{U}_{1,\epsilon},U_{2,\epsilon}}|}{|\mathbf{D}_{1}+\mathbf{\Sigma}_{\mathbf{U}_{1,\epsilon}}||d_2(\gamma)+\sigma^2_{U_{2,\epsilon}}|}\\
&\hspace{0.37in}+(N_2-N_1)\frac{\sigma^2_{S_2}+\sigma^2_{U_{2,\epsilon}}}{d_2(\gamma)+\sigma^2_{U_{2,\epsilon}}}-N_2\\
&=\lim\limits_{\epsilon\rightarrow 0}N_1\frac{|\mathbf{\Sigma}_{\mathbf{S}_1}+\mathbf{\Sigma}_{\mathbf{U}_{1,\epsilon}}|}{|\mathbf{D}_1+\mathbf{\Sigma}_{\mathbf{U}_{1,\epsilon}}|}-N_1.
\end{align*}
Therefore, the lower bound in (\ref{eq:separation}) is tight when $d_2>d_2(\sigma^2_{S_2})$, which completes the proof of Theorem \ref{thm:vectorscalar}.
\end{IEEEproof}

It is instructive to note that the role of $I(S_2+U_{2,\epsilon};Y_2)$ and $I(\mathbf{S}_1+\mathbf{U}_{1,\epsilon};\mathbf{S}_1(\gamma),S_2(\gamma)|S_2+U_{2,\epsilon})$ in the achievability argument is similar to that of $I(\mathbf{V}^n_2;Y^n_2)$ and $I(\mathbf{V}^n_1;Y^n_1|\mathbf{V}^n_2)$ in the proof of Corollary \ref{cor:lowersimplified}.
One can also readily see that (\ref{eq:useKKT3}) and (\ref{eq:diag}) imply
\begin{align*}
&\lim\limits_{\epsilon\rightarrow 0}I(\mathbf{S}_1+\mathbf{U}_{1,\epsilon},S_2+U_{2,\epsilon};\mathbf{S}_1(\gamma),S_2(\gamma))\\
&=I(\mathbf{S}_1,S_2;\mathbf{S}_1(\gamma),S_2(\gamma)),\\
&I(\mathbf{S}_1+\mathbf{U}_{1,\epsilon},S_2+U_{2,\epsilon}|\mathbf{S}_1(\gamma),S_2(\gamma))=0,
\end{align*}
respectively. These two equations can be viewed as the counterparts of (\ref{eq:para1}) and (\ref{eq:para2}).

It is implicitly assumed in our construction that $\mathbf{\Sigma}_{\mathbf{S}_1,S_2}\succ\mathbf{0}$, $\mathbf{D}_1\succ\mathbf{0}$, and $d_2>0$. In fact, Theorem \ref{thm:vectorscalar} also holds in the degenerate case where the source covariance matrix and the distortions are not strictly positive definite, i.e., we can relax the condition to  $\mathbf{\Sigma}_{\mathbf{S}_1,S_2}\succeq\mathbf{0}$ (which includes the case where $S_2$ is a linear function of $\mathbf{S}_1$), $\mathbf{D}_1\succeq\mathbf{0}$, and $d_2\geq0$. It is straightforward to verify that Corollary \ref{cor:lowersimplified} is directly applicable in this setup. For the achievability part, one can leverage the construction for the non-degenerate case via a simple perturbation argument. The details are left to the interested reader.

\subsection{Alternative Optimal Hybrid Schemes}\label{sec:optimal}

It turns out that in the vector-scalar case the hybrid scheme that achieves the optimal tradeoff between the transmit  power and the reconstruction distortion pair is in general not unique. Specifically, we shall show that if the optimal solution to (\ref{eq:opt}) is of the form\footnote{Note that this condition is satisfied if $\mbox{diag}(\mathbf{D}_1,\gamma)\preceq\mathbf{\Sigma}_{\mathbf{S}_1,S_2}$. In this case it follows by Fischer's inequality that $\mathbf{\Theta}(\gamma)=\mbox{diag}(\mathbf{D}_1,\gamma)$.} $\mathbf{\Theta}(\gamma)=\mbox{diag}(\mathbf{\Theta}_1(\gamma),\theta_2(\gamma))$, then there exists a class of hybrid schemes with the same performance as that in Section \ref{subsec:upper}.

Some additional notation needs to be introduced first. Recall $\mathbf{S}_1(\gamma)$, $S_2(\gamma)$, $\mathbf{Q}_1$, and $Q_2$ defined in Section \ref{subsec:upper}, and define $\Delta=S_2(\gamma)-\mathbb{E}[S_2(\gamma)|\mathbf{S}_1(\gamma)]$. Now write $\Delta=\Delta_0+\Delta_1+\Delta_2$, where $\Delta_0$, $\Delta_1$, and $\Delta_2$ are mutually independent zero-mean Gaussian random variables with variances to be specified. Furthermore, let $S_0(\gamma)=\mathbb{E}[S_2(\gamma)|\mathbf{S}_1(\gamma)]+\Delta_0$ and $S'_2(\gamma)=S_0(\gamma)+\Delta_1$. Note that $(\mathbf{Q}_1,Q_2)$ is independent of $(\Delta_0,\Delta_1,\Delta_2)$; moreover, since $\mathbf{\Theta}(\gamma)=\mbox{diag}(\mathbf{\Theta}_1(\gamma),\theta_2(\gamma))$, it follows that $\mathbf{Q}_1$ and $Q_2$ are mutually independent. Therefore, $\mathbf{S}_1\leftrightarrow \mathbf{S}_1(\gamma)\leftrightarrow S_0(\gamma)\leftrightarrow S'_2(\gamma)\leftrightarrow S_2$ form a Markov chain.

Note that
\begin{align*}
S_0(\gamma)&=\mathbb{E}[S_0(\gamma)|\mathbf{S}_1,S_2]+\bar{W}_0\\
&=\bar{\mathbf{a}}^T_1\mathbf{S}_1+\bar{a}_2S_2+\bar{W}_0,\\
\mathbf{S}_1(\gamma)&=\mathbb{E}[\mathbf{S}_1(\gamma)|\mathbf{S}_1,S_0(\gamma)]+\bar{\mathbf{W}}_1\\
&=\bar{\mathbf{B}}_1\mathbf{S}_1+\bar{\mathbf{b}}_2S_0(\gamma)+\bar{\mathbf{W}}_1,\\
S'_2(\gamma)&=\mathbb{E}[S'_2(\gamma)|S_2,S_0(\gamma)]+\bar{W}_2\\
&=\bar{c}_1S_2+\bar{c}_2S_0(\gamma)+\bar{W}_2,
\end{align*}
where $\bar{W}_0$ is independent of $(\mathbf{S}_1,S_2)$, $\bar{\mathbf{W}}_1$ is independent of $(\mathbf{S}_1,S_2,S_0(\gamma),S'_2(\gamma))$, and $\bar{W}_2$ is independent of $(\mathbf{S}_1,S_2,S_0(\gamma),\mathbf{S}_1(\gamma))$. We define
\begin{align*}
&\bar{\mathbf{S}}_1(\gamma)= \bar{\mathbf{B}}_1\mathbf{S}_1+\bar{\mathbf{W}}_1,\\
&\bar{S}_2(\gamma)=\bar{c}_1S_2+\bar{W}_2.
\end{align*}

\begin{figure*}[tb]
\begin{centering}
\includegraphics[width=15cm]{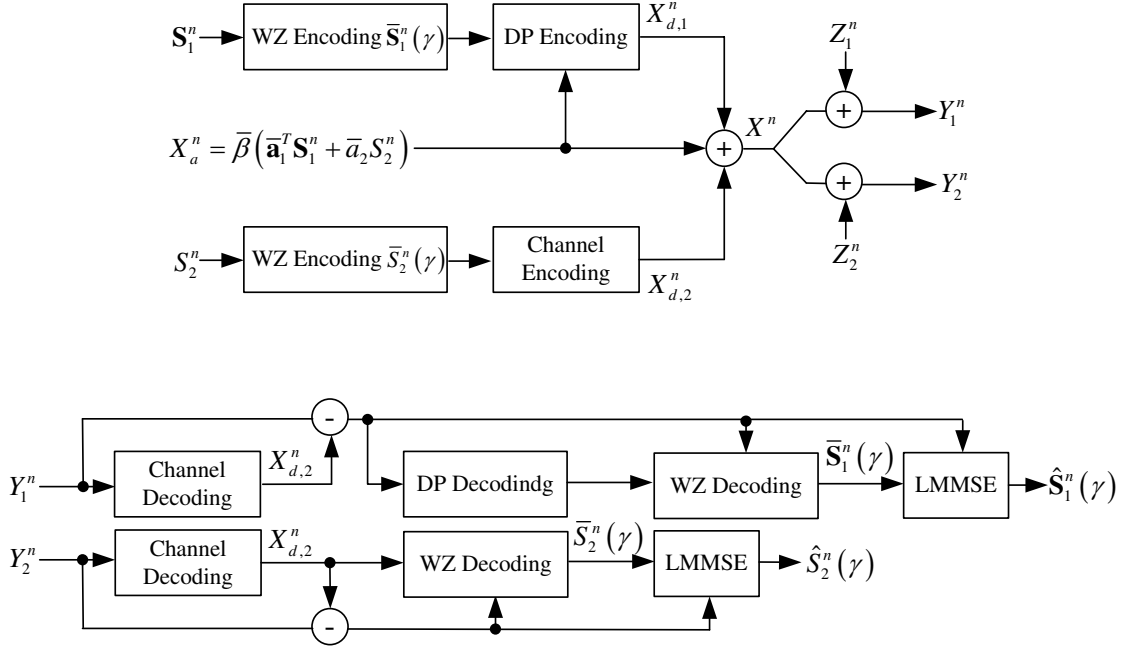}
\caption{An illustration of the hybrid scheme in Section \ref{sec:optimal}.\label{fig:hybrid2}}
\end{centering}
\end{figure*}

We are now in a position to describe the scheme (See Fig. \ref{fig:hybrid2}).

\vspace{0.5cm}
\noindent\textbf{Encoding:}
Let the channel input $X^n$, with average power $P(\gamma)$, be a superposition of an analog signal $X^n_a$ and two digital signals $X^n_{d,1}$ and $X^n_{d,2}$ (i.e., $X^n=X^n_a+X^n_{d,1}+X^n_{d,2}$).  The analog portion is given by $X^n_a=\bar{\beta}(\bar{\mathbf{a}}^T_1\mathbf{S}^n_1+\bar{a}_2S^n_2)$ for some non-negative number $\bar{\beta}$ to be specified later. For the digital portion $X^n_{d,2}$, the encoder first uses a Wyner-Ziv code  of rate $R_2$ with codewords  generated according to $\bar{S}_2(\gamma)$, with $S^n_2$ as the input, and with $X^n_a+X^n_{d,1}+Z^n_2$ as the decoder side information; the encoder then determines $X^n_{d,2}$ to send the bin index of the chosen Wyner-Ziv codeword $\bar{S}^n_2(\gamma)$ by using a channel code of rate $R_2$.
For the digital portion $X^n_{d,1}$, the encoder first uses a Wyner-Ziv code  of rate $R_1$ with codewords  generated according to $\bar{\mathbf{S}}_1(\gamma)$, with $\mathbf{S}^n_1$ as the input, and with $X^n_a+X^n_{d,1}+Z^n_1$ as the decoder side information; the encoder then determines $X^n_{d,1}$ to send the bin index of the chosen Wyner-Ziv codeword $\bar{\mathbf{S}}^n_1(\gamma)$ by using a dirty paper code of rate $R_1$ with $X^n_a$ treated as the channel state information known at the encoder. We define $P_a=\mathbb{E}[(X_a)^2]$ and $P_{d,i}=\mathbb{E}[(X_{d,i})^2]$, $i=1,2$, where $X_a\triangleq\bar{\beta}(\bar{\mathbf{a}}^T_1\mathbf{S}_1+\bar{a}_2S_2)$, $X_{d,1}$, $X_{d,2}$ are mutually independent zero-mean Gaussian random variables, and $P_a+P_{d,1}+P_{d,2}=P(\gamma)$.

\vspace{0.5cm}
\noindent\textbf{Decoding:} Receiver 2 decodes the channel code $X^n_{d,2}$, subtracts it from the channel output $Y^n_2\triangleq X^n_a+X^n_{d,1}+X^n_{d,2}+Z^n_2$, and recovers $\bar{S}^{n}_2(\gamma)$ by decoding the Wyner-Ziv code (the one of rate $R_2$) with $X^n_a+X^n_{d,1}+Z^n_2$ as the side information. Furthermore, in view of the fact
that the linear MMSE estimate of $S_2$  based on $(\bar{S}_2(\gamma), X_a+X_{d,1}+Z_2)$ is $\hat{S}_2(\gamma)\triangleq\rho_1\bar{S}_2(\gamma)+\rho_2(X_a+X_{d,1}+Z_2)$, where $(\rho_1,\rho_2)$ is an arbitrary solution to the following equation
\begin{align*}
&(\rho_1,\rho_2)\left(
                 \begin{array}{cc}
                   \mathbb{E}[(\bar{S}_2(\gamma))^2] & \mathbb{E}[\bar{S}_2(\gamma)X_a] \\
                   \mathbb{E}[\bar{S}_2(\gamma)X_a] & P_a+P_{d,1}+N_2 \\
                 \end{array}
               \right)\\
               &=(\mathbb{E}[S_2\bar{S}_2(\gamma)],\mathbb{E}[S_2X_a]),
 \end{align*}
 Receiver 2 can use $\hat{S}^n_2(\gamma)\triangleq\rho_1\bar{S}^n_2(\gamma)+\rho_2(X^n_a+X^n_{d,1}+Z^n_2)$ as the reconstruction of $S^n_2$; the resulting distortion is denoted by $d_2(\gamma)$. Receiver 1 also decodes the channel code $X^n_{d,2}$ and subtracts it from the channel output $Y^n_1\triangleq X^n_a+X^n_{d,1}+X^n_{d,2}+Z^n_1$. Then Receiver 1 decodes the dirty paper code and recovers $\bar{\mathbf{S}}^n_1(\gamma)$ by decoding the Wyner-Ziv code (the one of rate $R_1$) with $X^n_a+X^n_{d,1}+Z^n_1$ as the side information. Furthermore, in view of the fact that the linear MMSE estimate of $\mathbf{S}_1$ based on $(\bar{\mathbf{S}}_1(\gamma),X_a+X_{d,1}+Z_1)$ is $\hat{\mathbf{S}}_1(\gamma)\triangleq\bar{\mathbf{S}}_1(\gamma)+\bar{\beta}^{-1}\bar{\mathbf{b}}_2(X_a+X_{d,1}+Z_1)$,
  Receiver 1 can use $\hat{\mathbf{S}}^n_1(\gamma)\triangleq\bar{\mathbf{S}}^n_1(\gamma)+\bar{\beta}^{-1}\bar{\mathbf{b}}_2(X^n_a+X^n_{d,1}+Z^n_1)$ as the reconstruction of $\mathbf{S}^n_1$.

\vspace{0.5cm}
\noindent\textbf{Coding Parameters:}
Seven parameters $\mathbb{E}[(\Delta_0)^2]$, $\mathbb{E}[(\Delta_1)^2]$, $\bar{\beta}$, $R_1$, $R_2$, $P_{d,1}$, and $P_{d_2}$ still need to specified. Equivalently, we shall specify $\mathbb{E}[(\Delta_0)^2]$, $\mathbb{E}[(\Delta_1)^2]$, $P_a$, $R_1$, $R_2$, $P(\gamma)$, and $P_{d_2}$.

We again choose $P(\gamma)$ such that
\begin{align}
 I(\mathbf{S}_1,S_2;\mathbf{S}_1(\gamma),S_2(\gamma))=\frac{1}{2}\log\frac{P(\gamma)+N_1}{N_1}.\label{eq:poweragain}
\end{align}
Let $P_{d,2}$ be an arbitrary number in $[0,P^*_{d,2}]$, where $P^*_{d,2}$ is determined by the following equation
\begin{align*}
\frac{1}{2}\log\frac{P(\gamma)-P^*_{d,2}+N_1}{N_1}=I(\mathbf{S}_1,S_2;\mathbf{S}_1(\gamma)).
\end{align*}
Note that $P^*_{d,2}$ is nonnegative since
\begin{align*}
I(\mathbf{S}_1,S_2;\mathbf{S}_1(\gamma))&\leq I(\mathbf{S}_1,S_2;\mathbf{S}_1(\gamma),S_2(\gamma))\\
&=\frac{1}{2}\log\frac{P(\gamma)+N_1}{N_1}.
\end{align*}
Now choose $\mathbb{E}[(\Delta_0)^2]$ such that
\begin{align}
I(\mathbf{S}_1,S_2;\mathbf{S}_1(\gamma),S_0(\gamma))=\frac{1}{2}\log\frac{P(\gamma)-P_{d,2}+N_1}{N_1}.\label{eq:delta2}
\end{align}
The existence of such $\mathbb{E}[(\Delta_0)^2]$ is guaranteed by the fact that one can let $I(\mathbf{S}_1,S_2;\mathbf{S}_1(\gamma),S_0(\gamma))$ take any value in $[I(\mathbf{S}_1,S_2;\mathbf{S}_1(\gamma)),I(\mathbf{S}_1,S_2;\mathbf{S}_1(\gamma),S_2(\gamma))]$ (i.e., $[\frac{1}{2}\log\frac{P(\gamma)-P^*_{d,2}+N_1}{N_1},\frac{1}{2}\log\frac{P(\gamma)+N_1}{N_1}]$) by varying $\mathbb{E}[(\Delta_0)^2]$. We then choose $P_a\in[0,P(\gamma)-P_{d,2}]$ (which further determines $P_{d,1}$ and $\bar{\beta}$) such that
\begin{align}
I(X_a;X_a+X_{d,1}+Z_1)=I(\mathbf{S}_1,S_2;S_0(\gamma)), \label{eq:statisticequi2}
\end{align}
which is always possible in view of (\ref{eq:delta2}) and the fact that one can let $I(X_a;X_a+X_{d,1}+Z_1)$ take any value in $[0,\frac{1}{2}\log\frac{P(\gamma)-P_{d,2}+N_1}{N_1}]$ by varying $P_a$.
Next we set
\begin{align}
R_1=I(\mathbf{S}_1;\bar{\mathbf{S}}_1(\gamma)|X_a+X_{d,1}+Z_1).\label{eq:WZ1}
\end{align}
We finally choose $\mathbb{E}[(\Delta_1)^2]$ such that
\begin{align}
&I(S_2;\bar{S}_2(\gamma)|X_a+X_{d,1}+Z_2)\nonumber\\
&=\frac{1}{2}\log\frac{P(\gamma)+N_2}{P_a+P_{d,1}+N_2}\label{eq:channelcode}
\end{align}
and set
\begin{align}
R_2=I(S_2;\bar{S}_2(\gamma)|X_a+X_{d,1}+Z_2).\label{eq:WZ2}
\end{align}

It is not immediately clear that our particular choice of $\mathbb{E}[(\Delta_1)^2]$ always exists. To stress the dependence of $I(S_2;\bar{S}_2(\gamma)|X_a+X_{d,1}+Z_2)$ on $\mathbb{E}[(\Delta_1)^2]$, we shall denote it by $\psi(\mathbb{E}[(\Delta_1)^2])$.
Note that (\ref{eq:statisticequi2}), together with the fact that $I(X_a;X_a+X_{d,1}+Z_1)=I(\mathbf{S}_1,S_2;X_a+X_{d,1}+Z_1)$, implies that $I(\mathbf{S}_1,S_2;X_a+X_{d,1}+Z_1)=I(\mathbf{S}_1,S_2;S_0(\gamma))$; moreover, since both $X_{d,1}+Z_1$ and $\bar{W}_0$, which are Gaussian random variables, are independent of $(\mathbf{S}_1,S_2)$, it follows that the joint distributions of $(\mathbf{S}_1,S_2,\bar{\beta}^{-1}(X_a+X_{d,1}+Z_1))$ and $(\mathbf{S}_1,S_2,S_0(\gamma))$ are identical, which, in view of the fact that $\bar{\mathbf{W}}_1$ and $\bar{W}_2$ are independent of $(\mathbf{S}_1,S_2,S_0(\gamma),X_a+X_{d,1}+Z_1)$, further implies that the joint distributions of $(\mathbf{S}_1,S_2,\bar{\beta}^{-1}(X_a+X_{d,1}+Z_1),\bar{\mathbf{S}}_1(\gamma),\bar{S}_2(\gamma))$ and $(\mathbf{S}_1,S_2,S_0(\gamma),\bar{\mathbf{S}}_1(\gamma),\bar{S}_2(\gamma))$ are identical\footnote{We
   have implicitly assumed that $\mathbb{E}[(\bar{\mathbf{a}}^T_1\mathbf{S}_1+\bar{a}_2S_2)^2]>0$ (which implies that the $P_a$ and the $\bar{\beta}$ determined by (\ref{eq:statisticequi2}) are positive). For the degenerate case $\bar{\mathbf{a}}^T_1\mathbf{S}_1+\bar{a}_2S_2=0$ (which is possible if and only if $S_0(\gamma)=0$), one can simply set $X_a=0$ and $\bar{\beta}^{-1}(X_a+X_{d,1}+Z_1)=0$.}. Therefore, we have
\begin{align}
&\psi(\mathbb{E}[(\Delta_1)^2])\nonumber\\
&=I(X_a+X_{d,1}+Z_2,S_2;\bar{S}_2(\gamma))\nonumber\\
&\quad-I(X_a+X_{d,1}+Z_2;\bar{S}_2(\gamma))\nonumber\\
&=I(S_2;\bar{S}_2(\gamma))-I(X_a+X_{d,1}+Z_2;\bar{S}_2(\gamma))\label{eq:duetoMarkov}\\
&\geq I(S_2;\bar{S}_2(\gamma))-I(X_a+X_{d,1}+Z_1;\bar{S}_2(\gamma))\nonumber\\
&=I(S_2;\bar{S}_2(\gamma))-I(S_0(\gamma);\bar{S}_2(\gamma))\nonumber\\
&=I(S_2;\bar{S}_2(\gamma)|S_0(\gamma))\nonumber\\
&=I(S_2;S'_2(\gamma)|S_0(\gamma)),\nonumber
\end{align}
where (\ref{eq:duetoMarkov}) is due to the fact that $(X_a+X_{d,1}+Z_2)\leftrightarrow S_2\leftrightarrow\bar{S}_2(\gamma)$ form a Markov chain.
Clearly, $\psi(\mathbb{E}[(\Delta_1)^2])$ is a continuous function of $\mathbb{E}[(\Delta_1)^2]$.
When $\mathbb{E}[(\Delta_1)^2]=0$, we have $S'_2(\gamma)=S_0(\gamma)$ (which implies $\bar{S}_2(\gamma)=0$) and consequently $\psi(0)=0$; when $\mathbb{E}[(\Delta_1)^2]=\mathbb{E}[(\Delta)^2]-\mathbb{E}[(\Delta_0)^2]$, we have $S'_2(\gamma)=S_2(\gamma)$ and consequently
$\psi(\mathbb{E}[(\Delta)^2]-\mathbb{E}[(\Delta_0)^2])\geq I(S_2;S_2(\gamma)|S_0(\gamma))$. Note that
\begin{align}
&\frac{1}{2}\log\frac{P(\gamma)+N_1}{N_1}\nonumber\\
&=I(\mathbf{S}_1,S_2;\mathbf{S}_1(\gamma),S_2(\gamma))\label{eq:useinfo}\\
&=I(\mathbf{S}_1,S_2;\mathbf{S}_1(\gamma),S_2(\gamma),S_0(\gamma))\nonumber\\
&=I(\mathbf{S}_1,S_2;\mathbf{S}_1(\gamma),S_0(\gamma))+I(S_2;S_2(\gamma)|S_0(\gamma))\nonumber\\
&=\frac{1}{2}\log\frac{P_a+P_{d,1}+N_1}{N_1}+I(S_2;S_2(\gamma)|S_0(\gamma)),\label{eq:usestat}
\end{align}
where (\ref{eq:useinfo}) and (\ref{eq:usestat}) are due to (\ref{eq:poweragain}) and (\ref{eq:delta2}), respectively. This implies
\begin{align*}
I(S_2;S_2(\gamma)|S_0(\gamma))=\frac{1}{2}\log\frac{P(\gamma)+N_1}{P_a+P_{d,1}+N_1}.
\end{align*}
Therefore, we have
\begin{align*}
\psi(\mathbb{E}[\Delta^2]-\mathbb{E}[\Delta^2_0])&\geq\frac{1}{2}\log\frac{P(\gamma)+N_1}{P_a+P_{d,1}+N_1}\\
&\geq\frac{1}{2}\log\frac{P(\gamma)+N_2}{P_a+P_{d,1}+N_2}.
\end{align*}
Hence, our choice of $\mathbb{E}[\Delta^2_1]$ indeed exists.

\vspace{0.5cm}
\noindent\textbf{Conditions for Correct Decoding:} Receiver 2 needs to decode the channel code and the corresponding Wyner-Ziv code of rate $R_2$, and the correct decoding of these two components are guaranteed by  (\ref{eq:channelcode}) and (\ref{eq:WZ2}). Since Receiver 1 is stronger than Receiver 2, it can also decode the channel code and subtract it from the channel output. Receiver 1 additionally needs to decode the dirty paper code and the corresponding Wyner-Ziv code of rate $R_1$, the latter of which is guaranteed by (\ref{eq:WZ1}).

Recall that the joint distributions of $(\mathbf{S}_1,S_2,\bar{\beta}^{-1}(X_a+X_{d,1}+Z_1),\bar{\mathbf{S}}_1(\gamma),\bar{S}_2(\gamma))$ and $(\mathbf{S}_1,S_2,S_0(\gamma),\bar{\mathbf{S}}_1(\gamma),\bar{S}_2(\gamma))$ are identical.
Therefore, we have
\begin{align}
R_1&=I(\mathbf{S}_1;\bar{\mathbf{S}}_1(\gamma)|X_a+X_{d,1}+Z_1)\nonumber\\
&=I(\mathbf{S}_1;\mathbf{S}_1(\gamma)|S_0(\gamma))\nonumber\\
&=I(\mathbf{S}_1,S_2;\mathbf{S}_1(\gamma)|S_0(\gamma))\label{eq:inserts2}\\
&=I(\mathbf{S}_1,S_2;\mathbf{S}_1(\gamma),S_0(\gamma))-I(\mathbf{S}_1,S_2;S_0(\gamma))\nonumber\\
&=\frac{1}{2}\log\frac{P_a+P_{d,1}+N_1}{N_1}-\frac{1}{2}\log\frac{P_a+P_{d,1}+N_1}{P_{d,1}+N_1}\label{eq:plugin}\\
&=\frac{1}{2}\log\frac{P_{d,1}+N_1}{N_1},\label{eq:R1touse}
\end{align}
where (\ref{eq:inserts2}) follows by the fact that $S_2\leftrightarrow(S_0(\gamma),\mathbf{S}_1)\leftrightarrow\mathbf{S}_1(\gamma)$ form a Markov chain (which is implied by the fact that $S_2-S_0(\gamma)\leftrightarrow\mathbf{S}_1(\gamma)\leftrightarrow\mathbf{S}_1$ form a Markov chain), and (\ref{eq:plugin}) is due to (\ref{eq:delta2}) and (\ref{eq:statisticequi2}). Thus indeed Receiver 1 can decode the dirty paper code correctly.

\vspace{0.5cm}
\noindent\textbf{Optimality of this Class of Schemes:} Since the joint distributions of $(\mathbf{S}_1,\hat{\mathbf{S}}_1(\gamma))$ and $(\mathbf{S}_1,\mathbf{S}_1(\gamma))$ are identical (which is due to the fact that the joint distributions of $(\mathbf{S}_1,\bar{\beta}^{-1}(X_a+X_{d,1}+Z_1),\bar{\mathbf{S}}_1(\gamma))$ and $(\mathbf{S}_1,S_0(\gamma),\bar{\mathbf{S}}_1(\gamma))$ are identical), it follows that the resulting distortion at Receiver 1 is $\mathbf{\Theta}_1(\gamma)$, which is the same as that achieved by the optimal scheme given in Section \ref{subsec:upper}. We next focus on the distortion achieved at Receiver 2.

\begin{figure*}[tb]
\begin{centering}
\includegraphics[width=15cm]{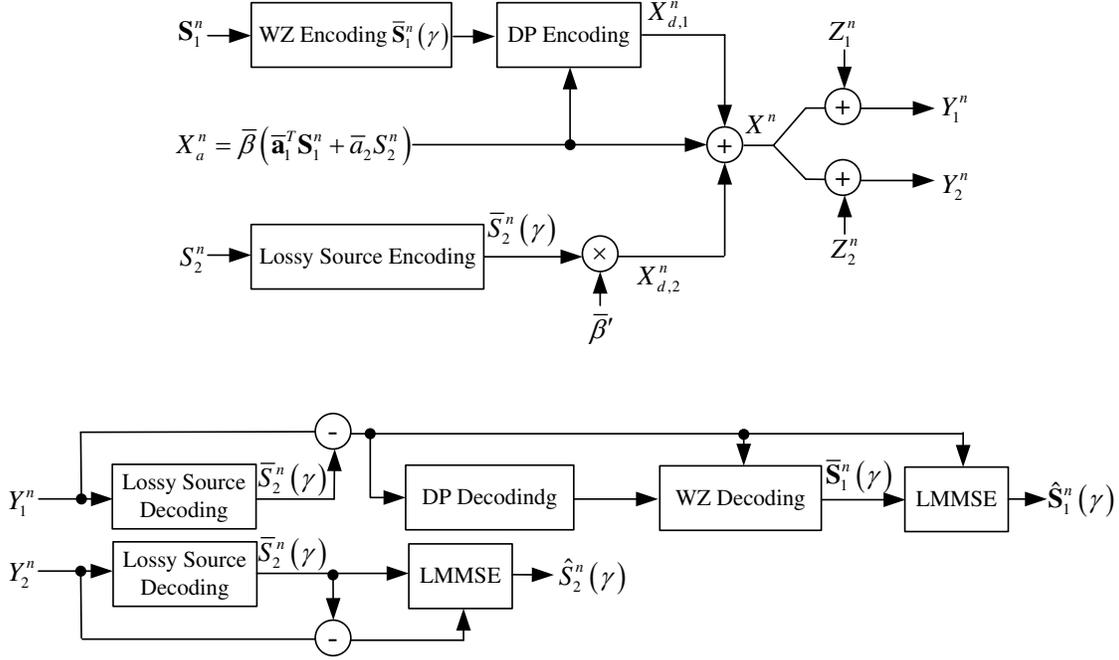}
\caption{A variant of the hybrid scheme in Section \ref{sec:optimal}.\label{fig:hybrid3}}
\end{centering}
\end{figure*}

Note that we have the freedom to choose $P_{d,2}$ from $[0,P^*_{d,2}]$. In particular, one can recover the hybrid scheme in Section \ref{subsec:upper} by setting $P_{d,2}=0$. We shall show\footnote{It is clear that the reconstruction distortion at Receiver 1 (i.e., $\mathbf{\Theta}_1(\gamma)$) does not depend on $P_{d,2}$} that the reconstruction distortion at Receiver 2 (i.e., $d_2(\gamma)$) does not depend on $P_{d,2}$; as a consequence, this class of schemes have exactly the same performance, and can all achieve the optimal tradeoff between the transmit power and the reconstruction distortion pair. Note that
\begin{align}
&\frac{1}{2}\log\frac{|\mathbb{E}[(\mathbf{S}_1-\mathbb{E}[\mathbf{S}_1|S_2])(\mathbf{S}_1-\mathbb{E}[\mathbf{S}_1|S_2])^T]|}{|\mathbf{\Theta}_1(\gamma)|}\nonumber\\
&=h(\mathbf{S}_1|S_2)-h(\mathbf{S}_1|\mathbf{S}_1(\gamma))\nonumber\\
&=h(\mathbf{S}_1|S_2)-h(\mathbf{S}_1|S_2,S_0(\gamma),\mathbf{S}_1(\gamma))\label{eq:toexp1}\\
&=I(\mathbf{S}_1;S_0(\gamma),\mathbf{S}_1(\gamma)|S_2)\nonumber\\
&=I(\mathbf{S}_1;X_a+X_{d,1}+Z_1,\hat{\mathbf{S}}_1(\gamma)|S_2)\label{eq:suff}\\
&=I(\mathbf{S}_1;X_a+X_{d,1}+Z_1,\bar{\mathbf{S}}_1(\gamma)|S_2)\nonumber\\
&=I(\mathbf{S}_1;X_a+X_{d,1}+Z_1|S_2)\nonumber\\
&\quad+I(\mathbf{S}_1;\bar{\mathbf{S}}_1(\gamma)|X_a+X_{d,1}+Z_1)\nonumber\\
&=I(\mathbf{S}_1;X_a+X_{d,1}+Z_1|S_2)+\frac{1}{2}\log\frac{P_{d,1}+N_1}{N_1}\label{eq:R1touse2}\\
&=\frac{1}{2}\log\frac{\mathbb{E}[(X_a-\mathbb{E}[X_a|S_2])^2]+P_{d,1}+N_1}{P_{d,1}+N_1}\nonumber\\
&\quad+\frac{1}{2}\log\frac{P_{d,1}+N_1}{N_1}\nonumber\\
&=\frac{1}{2}\log\frac{\mathbb{E}[(X_a-\mathbb{E}[X_a|S_2])^2]+P_{d,1}+N_1}{N_1},\nonumber
\end{align}
where (\ref{eq:toexp1})  follows from the fact that $\mathbf{S}_1\leftrightarrow\mathbf{S}_1(\gamma)\leftrightarrow (S_2, S_0(\gamma))$ form a Markov chain, (\ref{eq:suff}) follows from the fact that the joint distributions of $(\mathbf{S}_1,S_2,\bar{\beta}^{-1}(X_a+X_{d,1}+Z_1),\hat{\mathbf{S}}_1(\gamma))$ and $(\mathbf{S}_1,S_2,S_0(\gamma),\mathbf{S}_1(\gamma))$ are identical, and (\ref{eq:R1touse2}) is due to (\ref{eq:R1touse}). Therefore, $\mathbb{E}[(X_a-\mathbb{E}[X_a|S_2])^2]+P_{d,1}$ is not affected by the choice of $P_{d,2}$. Since
\begin{align}
&\frac{1}{2}\log\frac{\sigma^2_{S_2}}{d_2(\gamma)}\nonumber\\
&=I(S_2;\bar{S}_2(\gamma), X_a+X_{d,1}+Z_2)\nonumber\\
&=I(S_2;X_a+X_{d,1}+Z_2)+I(S_2;\bar{S}_2(\gamma)|X_a+X_{d,1}+Z_2)\nonumber\\
&=I(S_2;X_a+X_{d,1}+Z_2)+\frac{1}{2}\log\frac{P(\gamma)+N_2}{P_a+P_{d,1}+N_2}\label{eq:R2}\\
&=\frac{1}{2}\log\frac{P_a+P_{d,1}+N_2}{\mathbb{E}[(X_a-\mathbb{E}[X_a|S_2])^2]+P_{d,1}+N_2}\nonumber\\
&\quad+\frac{1}{2}\log\frac{P(\gamma)+N_2}{P_a+P_{d,1}+N_2}\nonumber\\
&=\frac{1}{2}\log\frac{P(\gamma)+N_2}{\mathbb{E}[(X_a-\mathbb{E}[X_a|S_2])^2]+P_{d,1}+N_2},\label{eq:diffentr}
\end{align}
where (\ref{eq:R2}) is due to (\ref{eq:channelcode}), it follows that $d_2(\gamma)$ does not depend on $P_{d,2}$.

\vspace{0.5cm}
\noindent\textbf{A Variant of this Class of Optimal Schemes:} For each $P_{d,2}\in[0,P^*_{d,2}]$, the aforedescribed scheme has the following variant (see Fig. \ref{fig:hybrid3}). Now for the digital portion $X^n_{d,2}$, the encoder simply uses a lossy source code  of rate $I(S_2;\bar{S}_2(\gamma))$ with codewords generated according to $\bar{S}_2(\gamma)$ and with $S^n_2$ as the input, and sets
$X^n_{d,2}$ to be the output codeword $\bar{S}^n_2(\gamma)$  multiplied by some non-negative number $\bar{\beta}'$, where $\bar{\beta}'$ is chosen such that $\mathbb{E}[(X_a+X_{d,1}+\bar{\beta}'\bar{S}_2(\gamma))^2]=P(\gamma)$. The remaining part of the encoder is still the same. Define $Y_i=X_a+X_{d,1}+\bar{\beta}'\bar{S}_2(\gamma)+Z_i$, $i=1,2$. Note that
\begin{align}
&I(S_2;\bar{S}_2(\gamma))+I(S_2;X_a+X_{d,1}+Z_2|\bar{S}_2(\gamma))\nonumber\\
&=I(S_2;\bar{S}_2(\gamma),X_a+X_{d,1}+Z_2)\nonumber\\
&=\frac{1}{2}\log\frac{\sigma^2_{S_2}}{d_2(\gamma)}\nonumber\\
&=\frac{1}{2}\log\frac{P(\gamma)+N_2}{\mathbb{E}[(X_a-\mathbb{E}[X_a|S_2])^2]+P_{d,1}+N_2}\label{eq:dist}\\
&=h(Y_2)-h(Y_2|S_2,\bar{S}_2(\gamma))\nonumber\\
&=I(S_2,\bar{S}_2(\gamma);Y_2)\nonumber\\
&=I(\bar{S}_2(\gamma);Y_2)+I(S_2;Y_2|\bar{S}_2(\gamma))\nonumber\\
&=I(\bar{S}_2(\gamma);Y_2)+I(S_2;X_a+X_{d,1}+Z_2|\bar{S}_2(\gamma)),\nonumber
\end{align}
where (\ref{eq:dist}) is due to (\ref{eq:diffentr}). This implies
\begin{align*}
I(S_2;\bar{S}_2(\gamma))=I(\bar{S}_2(\gamma);Y_2).
\end{align*}
Hence, Receiver 2 can decode the lossy source code and recover $\bar{S}^n_2(\gamma)$. Furthermore, Receiver 2 can\footnote{Note that Receiver 2 can obtain $X^n_a+X^n_{d,1}+Z^n_2$ from $\bar{S}^n_2(\gamma)$ and $Y^n_2\triangleq X^n_a+X^n_{d,1}+\bar{\beta}'\bar{S}^n_2(\gamma)+Z^n_2$.} use $\hat{S}^n_2(\gamma)$ as the reconstruction of $S^n_2$, and the resulting distortion is $d_2(\gamma)$. Receiver 1 can also decode the lossy source code and obtain $X^n_a+X^n_{d,1}+Z^n_1$ based on $\bar{S}^n_2(\gamma)$ and $Y^n_1\triangleq X^n_a+X^n_{d,1}+\bar{\beta}'\bar{S}^n_2(\gamma)+Z^n_1$.
Then Receiver 1 decodes the dirty paper code and recovers $\bar{\mathbf{S}}^n_1(\gamma)$ by decoding the Wyner-Ziv code (the one of rate $R_1$) with $X^n_a+X^n_{d,1}+Z^n_1$ as the side information. Moreover, Receiver 1 can use $\hat{\mathbf{S}}^n_1(\gamma)$ as the reconstruction of $\mathbf{S}^n_1$, and the resulting distortion is $\mathbf{\Theta}_1(\gamma)$. Therefore, this scheme has exactly the same performance as the original one. It is worth mentioning that the scheme in \cite{TDS11} can be viewed as an extremal case of this scheme with $P_{d,2}=P^*_{d,2}$ and $m_1=1$.

\section{Conclusion}\label{sec:conclusion}

We have obtained a lower bound on the optimal tradeoff between the transmit  power and the achievable distortion pair for the problem of sending correlated vector Gaussian sources over a Gaussian broadcast channel, where each receiver wishes to reconstruct its target source under a covariance distortion constraint. This lower bound is shown to be achievable by a class of hybrid schemes for the vector-scalar case, i.e., the scenario where the weak receiver wishes to reconstruct a  scalar source under the mean squared error distortion constraint. For certain classes of sources and distortion matrices, it is possible to extend our hybrid schemes to obtain a characterization of the optimal power-distortion tradeoff for the case where the weak receiver also wishes to reconstruct a vector source. However, a complete solution for this general setup remains elusive.

\appendices

\section{Proof of Lemma \ref{lem:entropybound}}\label{app:entropybound}

Let $\mathbf{W}(t)$ and $\hat{\mathbf{W}}(t)$ be the $t$-th columns of $\mathbf{W}$ and $\hat{\mathbf{W}}$, respectively, $t=1,\cdots,n$.  Note that
\begin{align*}
&h(\mathbf{W}|\hat{\mathbf{W}})\\
&=h(\mathbf{W}(1)|\hat{\mathbf{W}})+\sum\limits_{t=2}^nh(\mathbf{W}(t)|\hat{\mathbf{W}},\mathbf{W}(1),\cdots,\mathbf{W}(t-1))\\
&\leq\sum\limits_{t=1}^nh(\mathbf{W}(t)|\hat{\mathbf{W}}(t))\\
&\leq\sum\limits_{t=1}^nh(\mathbf{W}(t)-\hat{\mathbf{W}}(t))\\
&\leq\sum\limits_{t=1}^n\frac{1}{2}\log\left|2\pi e\mathbb{E}[(\mathbf{W}(t)-\hat{\mathbf{W}}(t))(\mathbf{W}(t)-\hat{\mathbf{W}}(t))^T]\right|\\
&\leq\frac{n}{2}\log\left|\frac{2\pi e}{n}\sum\limits_{t=1}^n\mathbb{E}[(\mathbf{W}(t)-\hat{\mathbf{W}}(t))(\mathbf{W}(t)-\hat{\mathbf{W}}(t))^T]\right|\\
&=\frac{n}{2}\log\left|\frac{2\pi e}{n}\mathbb{E}[(\mathbf{W}-\hat{\mathbf{W}})(\mathbf{W}-\hat{\mathbf{W}})^T]\right|,
\end{align*}
which completes the proof of Lemma \ref{lem:entropybound}.

\section{Proof of Lemma \ref{lem:condentropy}}\label{app:condentropy}

Let $\mathbf{W}_1(t)$ and $\mathbf{W}_2(t)$ be the $t$-th columns of $\mathbf{W}_1$ and $\mathbf{W}_2$, respectively, $t=1,\cdots,n$. Let $\Gamma$ be uniformly distributed over $\{1,\cdots,n\}$ and independent of $(\mathbf{W}_1,\mathbf{W}_2)$. We have
\begin{align}
&h(\mathbf{W}_1|\mathbf{W}_2)\nonumber\\
&=h(\mathbf{W}_1(1)|\mathbf{W}_2)\nonumber\\
&\quad+\sum\limits_{t=2}^nh(\mathbf{W}_1(t)|\mathbf{W}_2,\mathbf{W}_1(1),\cdots,\mathbf{W}_1(t-1))\nonumber\\
&\leq\sum\limits_{t=1}^nh(\mathbf{W}_1(t)|\mathbf{W}_2(t))\nonumber\\
&=nh(\mathbf{W}_1(\Gamma)|\mathbf{W}_2(\Gamma),\Gamma)\nonumber\\
&\leq nh(\mathbf{W}_1(\Gamma)|\mathbf{W}_2(\Gamma)).\label{eq:continue}
\end{align}
It is easy to see that
\begin{align*}
&\mathbb{E}[(\mathbf{W}^T_1(\Gamma),\mathbf{W}^T_2(\Gamma))^T(\mathbf{W}^T_1(\Gamma),\mathbf{W}^T_2(\Gamma))]\\
&=\frac{1}{n}\mathbb{E}[(\mathbf{W}^T_1,\mathbf{W}^T_2)^T(\mathbf{W}^T_1,\mathbf{W}^T_2)].
\end{align*}
Let $\hat{\mathbf{W}}_1(\Gamma)$ be the linear MMSE estimate of $\mathbf{W}_1(\Gamma)$  based on $\mathbf{W}_2(\Gamma)$. Note that
\begin{align}
&\left|\mathbb{E}((\mathbf{W}_1(\Gamma)-\hat{\mathbf{W}}_1(\Gamma))(\mathbf{W}_1(\Gamma)-\hat{\mathbf{W}}_1(\Gamma))^T)\right|\nonumber\\
&=\frac{\left|\mathbb{E}[(\mathbf{W}^T_1(\Gamma),\mathbf{W}^T_2(\Gamma))^T(\mathbf{W}^T_1(\Gamma),\mathbf{W}^T_2(\Gamma))]\right|}{\left|\mathbb{E}[\mathbf{W}_2(\Gamma)\mathbf{W}^T_2(\Gamma)]\right|}\nonumber\\
&=\frac{\left|\frac{1}{n}\mathbb{E}[(\mathbf{W}^T_1,\mathbf{W}^T_2)^T(\mathbf{W}^T_1,\mathbf{W}^T_2)]\right|}{\left|\frac{1}{n}\mathbb{E}[\mathbf{W}_2\mathbf{W}^T_2]\right|}.\label{eq:covariancetbu}
\end{align}
Now continuing from (\ref{eq:continue}),
\begin{align}
&nh(\mathbf{W}_1(\Gamma)|\mathbf{W}_2(\Gamma))\nonumber\\
&\leq nh(\mathbf{W}_1(\Gamma)-\hat{\mathbf{W}}_1(\Gamma))\nonumber\\
&\leq\frac{n}{2}\log\left|2\pi e\mathbb{E}((\mathbf{W}_1(\Gamma)-\hat{\mathbf{W}}_1(\Gamma))(\mathbf{W}_1(\Gamma)-\hat{\mathbf{W}}_1(\Gamma))^T)\right|\nonumber\\
&=\frac{n}{2}\log\frac{\left|\frac{2\pi e}{n}\mathbb{E}[(\mathbf{W}^T_1,\mathbf{W}^T_2)^T(\mathbf{W}^T_1,\mathbf{W}^T_2)]\right|}{\left|\frac{2\pi e}{n}\mathbb{E}[\mathbf{W}_2\mathbf{W}^T_2]\right|}, \label{eq:covariance}
\end{align}
where (\ref{eq:covariance}) is due to (\ref{eq:covariancetbu}). This completes the proof of Lemma \ref{lem:condentropy}.

\section{The Continuity of $\mathbf{\Theta}(\gamma)$}\label{app:theta}

     If $\mathbf{\Theta}(\gamma)$ is not continuous at $\gamma=\gamma^*$ for some $\gamma^*>0$,
     then there exists a sequence $\{\mathbf{\Theta}(\gamma_k)\}$ with $\gamma_k\rightarrow\gamma^*$ and $\mathbf{\Theta}(\gamma_k)\rightarrow\mathbf{\Theta}'(\gamma^*)\neq\mathbf{\Theta}(\gamma^*)$ as $k\rightarrow\infty$. Clearly, $\mathbf{\Theta}'(\gamma^*)$ satisfies the constraints for the maximization problem (with $\gamma=\gamma^*$) in (\ref{eq:opt}). Therefore, we must have $\log|\mathbf{\Theta}'(\gamma^*)|\leq\log|\mathbf{\Theta}(\gamma^*)|$. Now let $\tilde{\mathbf{\Theta}}(\gamma_k)=\mathbf{\Theta}(\gamma^*)-\mbox{diag}(\mathbf{0},\max(\gamma^*-\gamma_k,0))$. Note that $\tilde{\mathbf{\Theta}}(\gamma_k)$ satisfies the constraints for the maximization problem (with $\gamma=\gamma_k$) in (\ref{eq:opt}) when $\gamma_k$ is sufficiently close to $\gamma^*$. Therefore,
\begin{align*}
\limsup\limits_{k\rightarrow\infty}\log|\tilde{\mathbf{\Theta}}(\gamma_k)|\leq\lim\limits_{k\rightarrow\infty}\log|\mathbf{\Theta}(\gamma_k)|=\log|\mathbf{\Theta}'(\gamma^*)|.
\end{align*}
On the other hand, it is clear that
\begin{align*}
\lim\limits_{k\rightarrow\infty}\log|\tilde{\mathbf{\Theta}}(\gamma_k)|=\log|\mathbf{\Theta}(\gamma^*)|.
\end{align*}
Therefore, we must have $\log|\mathbf{\Theta}'(\gamma^*)|=\log|\mathbf{\Theta}(\gamma^*)|$, which, together with the uniqueness of $\mathbf{\Theta}(\gamma^*)$, implies $\mathbf{\Theta}'(\gamma^*)=\mathbf{\Theta}(\gamma^*)$. This leads to a contradiction.


\begin{thebibliography}{1}

\bibitem{Shannon:48}
C. E. Shannon,
\newblock ``A mathematical theory of communication,''
\newblock {\em Bell Syst. Tech. J.}, vol. 27, pp. 379-423, pp. 623--656, Jul., Oct. 1948.




\bibitem{MittalPhamdo:02}
U. Mittal and N. Phamdo,
\newblock ``Hybrid digital-analog (HDA) joint source-channel codes for broadcasting and robust communications,''
\newblock {\em IEEE Trans. Inf. Theory}, vol. 50, no. 5, pp. 1082--1102, May 2002.





\bibitem{PPR11}
V. M. Prabhakaran,  R. Puri, and K. Ramchandran, ``Hybrid digital-analog codes for source-channel
broadcast of Gaussian sources over Gaussian channels," \emph{IEEE Trans. Inf. Theory}, vol. 57, no. 7, pp. 4573--4588, Jul. 2011.

\bibitem{BAL11}
H. Behroozi, F. Alajaji, and T. Linder, ``On the performance of hybrid digital-analog coding
for broadcasting correlated Gaussian sources," {\em IEEE Trans. Commun.}, vol. 59, no. 12, pp. 3335--3342,
Dec. 2011.

\bibitem{KKE12}
A. Khina, Y. Kochman, and U. Erez, ``Joint unitary triangularization for MIMO networks," {\em IEEE Trans. Signal
Process.}, vol. 60, no. 1, pp. 326--336, Jan. 2012.

\bibitem{Lapidoth:09}
A. Lapidoth and S. Tinguely,
\newblock ``Sending a bivariate Gaussian over a Gaussian MAC,''
\newblock {\em IEEE Trans. Inf. Theory}, vol. 56, no. 6, pp.  2714--2752, Jun. 2010.



\bibitem{BLT10}
S. Bross, A. Lapidoth, and S. Tinguely, ``Broadcasting correlated Gaussians," {\em IEEE Trans. Inf. Theory}, vol.~56, no.~7, pp.~3057--3068, Jul.
2010.


\bibitem{SV09}
R. Soundararajan and S. Vishwanath, ``Hybrid coding for Gaussian broadcast channels with
Gaussian sources," in {\em Proc. IEEE Int. Symp. Inform. Theory (ISIT)}, Jun./Jul. 2009, pp. 2790--2794.

\bibitem{GT13}
Y. Gao and E. Tuncel, ``Separate source-channel coding for transmitting correlated Gaussian
sources over degraded broadcast channels," {\em IEEE Trans. Inf.
Theory}, vol. 59, no. 6, pp. 3619--3634, Jun. 2013.

\bibitem{TDS11}
C. Tian, S. Diggavi, and S. Shamai (Shitz), ``The achievable distortion region of
sending a bivariate Gaussian source on the
Gaussian broadcast channel," {\em IEEE Trans. Inf. Theory}, vol.~57, no.~10, pp.~6419--6427, Oct.
2011.






\bibitem{KSH00}
T. Kailath, A. H. Sayed, and B. Hassibi, {\em Linear Estimation}. Upper
Saddle River, NJ: Prentice-Hall, 2000.

\bibitem{RFZ06}
Z. Reznic, M. Feder, and R. Zamir, ``Distortion bounds for broadcasting with bandwidth expansion," {\em IEEE Trans. Inf. Theory},
vol. 52, no. 8, pp. 3778--3788, Aug. 2006.

\bibitem{KC14}
K. Khezeli and J. Chen, ``A source-channel separation theorem with application to the source broadcast problem," in {\em Proc. IEEE Int. Symp. Inform. Theory (ISIT)}, Honolulu, HI, USA, Jun./Jul. 2014, pp. 2132--2136.

\bibitem{Ozarow80}
L. Ozarow, ``On a source coding problem with two channels and
three receivers," {\em Bell Syst. Tech. J.}, vol.~59, no.~10,
pp.~1909--1921, Dec. 1980.

\bibitem{WV07}
H. Wang and P. Viswanath, ``Vector Gaussian multiple description
with individual and central receivers," {\em IEEE Trans. Inf.
Theory}, vol.~53, no. 6, pp.~2133--2153, Jun. 2007.

\bibitem{Chen09}
J. Chen, ``Rate region of Gaussian multiple description coding with individual and central distortion constraints," {\em IEEE Trans. Inf. Theory}, vol. 55, no. 9, pp. 3991--4005, Sep. 2009.

\bibitem{SSC13}
L. Song, S. Shao, and J. Chen, ``On the sum rate of multiple description coding with symmetric distortion constraints," {\em IEEE Trans. Inf. Theory}, vol. 60, no. 12, pp. 7547--7567, Dec. 2014.



\bibitem{PRP02}
R. Puri, K. Ramchandran, and S. S. Pradhan, ``On seamless digital upgrade of analog transmission systems using coding with side information," in {\em Proc. 40th Annu. Allerton Conf. Commun., Control, Comput. (Allerton)}, Monticello, IL, Oct. 2002.

\bibitem{PPR08}
V. M. Prabhakaran,  R. Puri, and K. Ramchandran, ``Colored Gaussian source–channel broadcast for
heterogeneous (analog/digital) receivers," \emph{IEEE Trans. Inf. Theory}, vol. 54, no. 4, pp. 1807–-1814, Apr. 2008.

\bibitem{WynerZiv:76}
A.~D. Wyner and J.~Ziv, ``The rate-distortion function for source coding with
  side information at the decoder,'' {\em IEEE Trans. Inf. Theory},
  vol. IT-22, no.~1, pp. 1--10, Jan. 1976.

\bibitem{Costa:83}
M. Costa,
\newblock ``Writing on dirty paper,''
\newblock {\em IEEE Trans. Inf. Theory}, vol. IT-29, no. 3, pp. 439--411, May 1983.


\bibitem{Bertsekas99}
D. P. Bertsekas, \emph{Nonlinear Programming}, 2nd ed. Belmont, MA: Athena Scientific, 1999.













\end{thebibliography}
\end{document}